# Versatile Top-Down Patterning Technique for Perovskite On-Chip Integration

*Federico Fabrizi[1,2,°], Saeed Goudarzi[2,°], Sana Khan[1,2], Tauheed Mohammad[1], Liudmila Starodubtceva[2], Piotr J. Cegielski[1], Felix Thiel[3], Sercan Özen[4], Maximilian Schiffer[5,6], Felix Lang[4], Peter Haring Bolívar[3], Thomas Riedl[5,6], Gerhard Müller-Newen[7], Surendra B. Anantharaman[1,8], Maryam Mohammadi[1,*], Max C. Lemme[1,2,*]*

[1] AMO GmbH, Otto-Blumenthal-Straße 25, Aachen, Germany.

[2] Chair of Electronic Devices, RWTH Aachen University, Otto-Blumenthal-Straße 25, Aachen, Germany.

[3] Institute for High Frequency and Quantum Electronics, University of Siegen, 57076 Siegen, Germany.

[4] Institute Freigeist Juniorgroup, Radiation Tolerant Electronics with Soft Semiconductors (ROSI), University of Potsdam, Karl-Liebknecht-Straße 24/25, Potsdam-Golm, Germany.

[5] Institute of Electronic Devices, University of Wuppertal, Rainer-Gruenter-Str. 21, 42119 Wuppertal, Germany.

[6]Wuppertal Center for Smart Materials & Systems (CM@S), University of Wuppertal, *Rainer-Gruenter-Str. 21,* 42119 Wuppertal, Germany

[7]Institute of Biochemistry and Molecular Biology, Uniklinik RWTH Aachen, Pauwelsstrasse 30, Aachen, Germany.

[8]Low-dimensional Semiconductors Lab, Department of Metallurgical and Materials Engineering, Indian Institute of Technology Madras, Chennai 600036, India.

°These authors contributed equally.

*Corresponding Authors – mohammadi@amo.de, max.lemme@eld.rwth-aachen.de

**Abstract**

Metal-halide perovskites (MHPs) have exciting optoelectronic properties and are under investigation for various applications, such as photovoltaics, light-emitting diodes, and lasers. An essential step towards exploiting the full potential of this class of materials is their large-scale, on-chip integration with high-resolution, top-down patterning. The development of such patterning methods for perovskite films is challenging because of their intrinsic ionic nature and adverse reactions with the solvents used in standard lithography processes. Here, we introduce a versatile and precise method comprising photolithography and reactive ion etching (RIE) processes that can be tuned to accommodate different perovskite compositions and morphologies. Our method utilizes conventional photoresists at reduced temperatures to create micron-sized features down to 1 µm, providing high reproducibility from chip to chip. The patterning technique is validated through atomic force microscopy (AFM), X-ray diffraction (XRD), optical spectroscopy, and scanning electron microscopy (SEM). It enables the scalable and high-throughput on-chip monolithic integration of MHPs.

## Introduction

Metal-halide perovskites (MHPs) are a class of ionic semiconductors known for their outstanding optoelectronic properties, including charge carrier diffusion lengths exceeding 1 µm[1–4] and tunable bandgap[5–7]. These properties have spawned substantial research in various fields, such as photovoltaics[8–10], photonics[11–14], and optoelectronics[15–19]. Furthermore, the optoelectronic properties of MHPs, including their bandgap, exciton binding energy, and charge carrier mobility, depend on both their composition and dimensionality[5,20]. The engineering of the composition of MHPs is facilitated by their rich chemistry, which provides opportunities for tuning their electronic and optical properties[21,22]. Additionally, the dimensionality of MHPs can be altered from three-dimensional (3D)[23] to two-dimensional (2D)[24], one-dimensional (1D)[25], and zero-dimensional (0D)[26].

Technologically, their simple, low-temperature solution-based deposition processes [27] are beneficial for upscaling production once the relevant technology readiness levels are achieved. However, integrated on-chip applications in optoelectronics and photonics, such as displays or light-emitting diodes (LEDs) [28–30], require compatibility with the perovskites' limited thermal budgets and complex semiconductor manufacturing processes [31–34]. One of the key processes for such integration is the patterning of perovskite materials[35,36]. The industry's standard and preferred technique combines top-down photolithography and reactive ion etching (RIE) because of its high resolution, precision, reliability, selectivity, and good cost-performance trade-off. However, traditional photolithography involves wet chemical processing in polar solvents and deionized water, which damages or dissolves highly ionic crystals such as perovskites[37,38]. This chemical instability and sensitivity to polar and protic solvents[39], which are present in photoresist spin-coating processes, lead to degradation or alteration of the intrinsic properties of the perovskite. Several alternative methods have been developed, such as nanoimprint patterning [40–43], fluorinated resists[44], gas-assisted focused ion beam etching[45], inkjet printing[46], laser writing[47,48] and wettability-assisted photolithography processes[49,50]. However, these methods do not generally meet the industry's needs for high-speed production on a wafer scale and accurate and consistent features.

Here, we demonstrate a precise and reproducible method for the top-down patterning of perovskite films with micrometer resolution based on standard semiconductor equipment and processes. The micrometer resolution remains highly relevant for a broad range of industrial applications, particularly in emerging technologies involving perovskite-based optoelectronics and hybrid semiconductor devices[51]. In perovskite solar cells[52], photodetectors[29,53],light-emitting devices[54], and lasers[28] feature sizes on the order of 1–10 µm are commonly used. Moreover, UV photolithography at micrometer scales is cost-effective, high-throughput, and compatible with soft perovskite thin films, avoiding possible damage associated with electron beam lithography (EBL) irradiation[55]. While alternative approaches such as nanoimprint lithography (NIL), focused ion beam (FIB), and laser writing can reach sub-micrometer or nanometer-scale resolution (see Table S1), these methods typically face significant challenges in terms of scalability, processing speed, and compatibility with perovskite materials. The presented method can be tailored for various perovskite types and morphologies, including 3D and low-dimensional (quasi-2D and 0D) structures. We use a double-stack resist composed of polymethyl methacrylate (PMMA) and AZ MIR 701, a commercial positive photoresist[56]. The PMMA prevents direct contact between the perovskite and photoresists[57], whereas AZ MIR 701 facilitates a standard lithography process. We subsequently employ RIE to etch the perovskites while avoiding oxygen plasma, which also damages the perovskites[39].

**Experiment**

Motivated by earlier works on perovskite-compatible photolithography methods [28,39,58], we devised a new strategy that integrates two resists and non-oxidative reactive gases into a singular workflow. In the following, we introduce the method and demonstrate its application to 3D cesium lead bromide ($CsPbBr_3$) and methylammonium lead iodide ($MAPbI_3$), quasi-2D ($PEA_2(MAPbBr_3)_{n-1}PbBr_4$) and 0D formamidinium lead bromide ($FAPbBr_3$) perovskites.

The different perovskite thin films were spin-coated onto a $Si/SiO_2$ substrate (see Experimental Section for more details). A 200 nm encapsulation layer (PMMA) was deposited on top of the perovskite layer via spin-coating at 4000 rpm for 60 seconds and baking at 80°C for 10 minutes. This layer acted as a spacer layer and protected the perovskite material from the photoresist. Then, the AZ MIR 701 photoresist was spin-coated on top of the samples at 3000 rpm for 60 s and soft-baked at 95°C for 90 s, after which the samples were exposed to ultraviolet (UV) light for 25 s via a contact lithography mask aligner (EVG 420) system. The patterns were developed in MF26A developer for 35 s followed by rinsing for 7 s with deionized water. Then, the samples were etched via a PlasmaLab System 100 inductively coupled plasma (ICP)-RIE tool (Oxford Instruments). All inorganic 3D metal halides, quasi-2D perovskites, and nanocrystalline thin films were etched with radicals from HBr and $BCl_3$ gases with different gas flows. The hybrid organic-inorganic perovskites were etched with radicals from $CF_4$, $Cl_2$ and He gas. Since the metal halide perovskites are sensitive to high temperatures, the etching processes were performed in a step and repeated fashion with one cycle of etching followed by a waiting cycle for two minutes to reduce the substrate temperatures. The etching rate of $CsPbBr_3$ was 100 nm/min, and the selectivity over $SiO_2$ was 3.6, whereas for $MAPbI_3$, the values were 90 nm/min and 2.7. The etch rate and selectivity over $SiO_2$ for the quasi-2D perovskite were 95 nm/min and 3.4, and the values for the 0D $FAPbBr_3$ were 92 nm/min and 2.8. Finally, the photoresist stack was removed by dissolving the PMMA encapsulation layer in the non-polar solvent toluene at 80°C for 1 hour to ensure complete dissolution of the PMMA. A schematic of the process is shown in Figure 1. The process yielded perovskite features as small as 1 µm, which is the smallest resolution achievable with our contact lithography system (Figures 1b and c).

## Results and Discussion

### 1. Photolithographic patterning of three-dimensional MHPs

We investigated our process with the 3D perovskites $CsPbBr_3$ and $MAPbI_3$ because they are among the most studied and technologically mature metal halide perovskites.

#### 1.1. All-inorganic perovskite ($CsPbBr_3$)

Figures 2a and b show atomic force microscopy (AFM) images of the surface topography of the $CsPbBr_3$ thin film before and after etching. The bright areas with greater heights are attributed to the material's secondary phases of $CsPb_2Br_5$ and $Cs_4PbBr_6$, as evidenced by X-ray diffraction (XRD) measurements (Figure 2c), corroborating previous findings[13]. The root mean square (RMS) roughness was 14.92 ± 1.57 nm, which increased to 24.09 ± 1.21 nm after the etching process. We also observed an increase in the average grain size from approximately 65 nm after spin coating to 200 nm after stripping the double-stack resist (Supporting Information (SI) Figures S1a and b) and an increase in the number of pinholes in the film. This suggests some degree of recrystallization during the etching process, despite tailoring the etching recipe to limit process-induced heating. It is important to note that derivative phases such as the two-dimensional $CsPb_2Br_5$ and the zero-dimensional $Cs_4PbBr_6$ can readily convert to the 3D $CsPbBr_3$ phase under post mild thermal treatment, due to their higher thermodynamic stability [59,60]. The AFM images in Figures 2a and 2b show a reduction in the bright contrast regions corresponding to secondary phases in the pristine film after patterning. The XRD data in Figure 2c show no significant differences in the crystalline phases before and after etching. The spectra show distinct high-intensity peaks at 15.21°, 21.49°, 26.37°, 30.69°, 34.46°, and 37.90°, corresponding to the (100), (110), (111), (200), ($\bar{2}10$), and ($\bar{2}11$) crystallographic planes, respectively, characteristic of monoclinic $CsPbBr_3$ (reference code: JCPDS 00-018-0364)[61,62]. Additional peaks were observed at 12.63°, 22.41°, 25.42°, 27.51°, and 28.63°, indicative of the presence of the (012), (300), (024), (131), and (214) crystallographic planes, respectively, associated with the rhombohedral $Cs_4PbBr_6$ phase[63]. Furthermore, a peak attributed to the (002) plane of the $CsPb_2Br_5$ phase was also observed, supporting the results of the AFM analysis[64,65]. The patterned perovskites were optically characterized through ultraviolet–visible (UV-Vis) absorption and photoluminescence (PL) spectroscopy. The absorption edges before and after the etching process were very similar, approximately 516 nm and 517 nm, respectively. The photoluminescence peaks remained at 521 nm before and after the etching process, with full widths at half maximum (FWHM) of 15.1 nm and 14.9 nm, respectively (see individual spectra in Figure 2d). PL emission maps taken over large scan areas of 25 µm × 25 µm confirmed this result. They are shown in Figures 2e and f, where the color scale indicates the peak position wavelength. The FWHM of the PL spectra also did not change before and after the etching process in the same scanning areas (SI Figures S1c and d). Moreover, no differences can be discerned in the PL spectra between the center and edges of the etched structures (SI Figure S1e). Time-resolved photoluminescence measurements were performed before and after etching, and the resulting decay curves were fitted using a double-exponential model (SI Figure S2). PL decay times $\tau_1$=1.39 ns and $\tau_2$=4.50 ns for the pristine film (SI Figure S2a and b), and $\tau_1$=1.07 ns and $\tau_2$=5.03 ns for the etched sample (SI Figure S2c and b), were extracted. These results show that etching slightly accelerates the fast decay component ($\tau_1$), likely due to increased surface recombination[66], while the slower component ($\tau_2$) remains relatively unchanged. This suggests that the optical properties of the material are preserved[66,67], supporting the potential for device applications after the patterning process. Photoluminescence quantum yield (PLQY) measurements show a slight enhancement from 1.02 % to 1.39 % after etching (SI Figure S3).

Figure 2g shows a cross-sectional SEM image of the etched $CsPbBr_3$ sidewall with an angle of approximately 85°. Furthermore, the top-view SEM image, profilometer measurement and optical microscope image after etching in SI Figures S1f, g and h display low edge roughness and well-defined structures. Confocal fluorescence spectra before and after etching, shown in SI Figure S1i, exhibits the material's characteristic emission in the green region of the spectrum. The confocal microscopy image of the pristine $CsPbBr_3$ thin film (SI Figure S1l) and the image after etching (Figure 2h), reveal uniform green fluorescence and display representative patterns that can be achieved through etching. To validate the method's precision, various features fabricated using the developed technique are shown in SI Figure S4. The applicability of our method to a broader range of all-inorganic perovskite materials, including those with wider band gaps, was confirmed by testing on cesium lead chloride ($CsPbCl_3$) (SI Figure S5).

### 1.2. Organic-inorganic hybrid perovskite ($MAPbI_3$)

We further investigated our etching method with the less stable organometallic halide perovskite $MAPbI_3$, following the same fabrication scheme illustrated in Figure 1.

The morphology of the $MAPbI_3$ thin film before and after etching was examined via AFM (Figures 3a and b). The RMS roughness slightly increased from 6.20 ± 1.99 nm to 7.06 ± 1.08 nm after etching. In contrast to $CsPbBr_3$, there were no noticeable changes in the grain size of $MAPbI_3$. (from 146.85 ± 70.94 nm to 158.64 ± 81.97 nm, SI Figures S6a and b).

The XRD spectra of $MAPbI_3$ before and after etching have dominant peaks at 14.09°, 19.93°, 23.54°, 24.52°, 28.45°, 31.85°, and 34.9°, corresponding to the (110), (200), (211), (202), (220), (310) and (312) crystallographic planes of the tetragonal phase of $MAPbI_3$, respectively (reference code: JCPDS No. 96-451-8044) [68–71]. A minor peak attributed to the $PbI_2$ phase is also evident at 12.7°[68]. After etching, an additional peak occurred at 15.1°, indicating a different material composition at the edges of the patterned areas. Confocal microscopy images of these areas, compared to the one of the pristine sample (SI Figure S6c), showed blue and green emissions all over the edges (SI Figures S6d, e, and f), which can be attributed to halide exchange with chlorine radicals from the etch plasma, resulting in the formation of $MAPbI_{3-x}Cl_x$ ($0<X\leq3$)[68]. This is enabled by the low crystal formation energy of organic-inorganic hybrid metal halides, facilitating the chloride ion diffusion. UV-Vis absorption and PL spectra of the patterned $MAPbI_3$ structures show absorption edges and photoluminescent peaks at 748 nm and 768 nm, respectively, and no significant wavelength shifts were observed before and after etching in the sample center (Figure 3d). This is also confirmed by the spectrum extracted from the confocal images (SI Figure S6g). The FWHM values of 37.4 nm and 37.2 nm before and after etching, respectively, are comparable. However, the region close to the edge of the $MAPbI_3$ shows a blueshift of approximately 7 nm in the PL peak position (SI Figure S6h). This can be explained by decreased average grain sizes at the edge of the etched structure (SI Figure S6i), which also contributed to an enhanced PL intensity (inset of SI Figure S6h). Similar effects were reported previously for perovskite thin films[28,73] and single crystals[74]. PL maps (Figures 3e and f) and an analysis of the FWHM (SI Figures S6l and m) over an area of 25 µm × 25 µm in the center of the samples demonstrate the optical stability of $MAPbI_3$ towards the patterning process. To further investigate the impact of the nanofabrication on the luminescence properties, we also conducted PLQY measurements of the pristine $MAPbI_3$ film and after the etching (SI Figure S7) and observed a reduction from 0.213 % to 0.172 %. This modest decrease indicates that our nanofabrication process preserves the optical quality of the $MAPbI_3$ film, despite its known sensitivity to processing.

Figure 3g shows a cross-sectional SEM image of the $MAPbI_3$ sidewall with an angle of approximately 80°. Moreover, top-view SEM image, optical microscope image, and profilometer measurement after etching (SI Figure S6n, o, and p) demonstrate the structured $MAPbI_3$ with a lateral size of 85 µm by 65 µm and a height of 250 nm. Figure 3h shows a confocal microscopy image of typical patterns that can be achieved with $MAPbI_3$, with the characteristic luminescence wavelength in the red range of the spectrum. A diverse set of patterns produced using the proposed technique is shown in SI Figure S8.

## 2. Photolithographic patterning of low-dimensional MHPs

### 2.1. Quasi-two-dimensional perovskites ($PEA_2(MAPbBr_3)_{n-1}PbBr_4$)

Next, we demonstrated the suitability of the patterning technique in Figure 1 for the fabrication of high-resolution quasi-2D perovskite arrays. AFM images in Figures 4a and b show that the morphology of the perovskites was retained after the process. The roughness of the film decreased from 57.51 ± 8.95 nm to 50.66 ± 15.84 nm, indicating a smoothing effect through the etching process. The XRD spectra in Figure 4c remained unchanged before and after patterning and show four sharp peaks at 5.07°, 14.95°, 21.38°, and 30.20°, corresponding to the (002), (100), (110), and (200) crystal planes of the quasi-2D perovskites, respectively[75]. The low-angle diffraction peak below 5° can be attributed to typical reflections from the layered structure [76]. The UV–Vis and PL spectra in Figure 4d show absorption edges and photoluminescence peaks at 517 nm and 528 nm, respectively, and no significant peak position shift before and after etching. This is uniform across the sample, as illustrated in the PL emission maps, with only a slight variability of 2 nm in the emission wavelength over an area of 25 µm x 25 µm (Figures 4e and f). An average reduction of approximately 2 nm in the FWHM was observed in area scans (SI Figures S9a and b). The PL spectra showed no significant differences in emission wavelengths between the center and the edge of the structures (SI

Figure S9c). However, PLQY measurements before and after the etching process (SI Figure S10) revealed a decrease from 2.11 % to 1.56 %. Time-resolved photoluminescence measurements showed a biexponential decay for the pristine film, with lifetimes $\tau_1$=3.23 ns and $\tau_2$=32.48 ns (SI Figure S11a and b), while the etched sample exhibited $\tau_1$=3.34 ns and $\tau_2$=47.53 ns (SI Figure S11c and b). The negligible change in $\tau_1$ after etching (from 3.23 ns to 3.34 ns) suggests that surface recombination or trap-related losses are not significantly affected by the etching process. In contrast, the notable increase in $\tau_2$ indicates a suppression of non-radiative bulk recombination, possibly due to enhanced passivation during the nanofabrication process.

The cross-sectional SEM image in Figure 4g shows well-defined structures after the etching process, with a sidewall angle of approximately 85°. The quality of the etched $PEA_2(MAPbBr_3)_{n-1}PbBr_4$ structures is shown in SI Figures S9d, e, and f. The top-view SEM image, optical microscope image, and profilometer measurement after etching illustrate the structured $PEA_2(MAPbBr_3)_{n-1}PbBr_4$ with a lateral size of 85 µm by 65 µm and a height of 290 nm. Confocal fluorescence spectra (SI Figure S9g) and the corresponding microscopy images (SI Figure S9h and Figure 4h) show uniform green emission before and after etching. Figure 4h confirms the successful etching of the quasi-2D perovskite with its green emission, with no residual perovskite in the etched area. To demonstrate the versatility of our process, we fabricated several shapes and patterns (SI Figure S12).

### 2.2. Zero-dimensional perovskites ($FAPbBr_3$)

Finally, we applied our micropatterning technique to thin films of colloidal $FAPbBr_3$ perovskite nanocrystals (NCs). The surface roughness of the NC layers was assessed by AFM before and after the etching process (Figures 5a and b), the film roughness decreased from 9.66 ± 0.79 nm to 2.03 ± 0.13 nm. The XRD spectra taken before and after patterning were not significantly affected (Figure 5c), with four characteristic peaks of the cubic structure of $FAPbBr_3$ that correspond to the (100), (110), (200) and (210) crystal planes (.cif file No. 7130785)[77,78]. The Scherrer equation ($D = \frac{K\lambda}{\beta cos\theta}$) can be used to measure the size of the nanocrystal. It revealed no significant reduction in the size: 9.66 nm and 9.12 nm before and after etching, respectively. The UV-Vis and PL spectra of the NCs exhibited differences in their absorption edges and photoluminescence peaks before and after the etching process. As illustrated in Figure 5d, both spectra show a blueshift of approximately 8 nm after the etching process. The absorption edge before the nanofabrication process was located at 518 nm, whereas it was 510 nm after the etching process. A similar trend was observed for the PL peak position, which moved from 521 nm to 513 nm. This is confirmed in Figures 5e and f, where an average blueshift of the PL peak position of 8 nm was extracted from the 25 µm X 25 µm area scan. The FWHM before and after etching remained very similar at 19.47 nm and 19.18 nm, respectively. This was confirmed in the respective area maps (SI Figures S13a and b). There were also no noticeable differences in the emission peak positions between the center and edge of the etched structure (SI Figure S13c). Time-resolved photoluminescence measurements revealed similar values for $\tau_1$ before and after etching ($\tau_1$=2.63 ns to 2.49 ns), while showing an increase for $\tau_2$ after etching ($\tau_2$=8.65 ns to 11.37 ns) (SI Figure S14). PLQY decreased from 51.8 % to 11.7 % (SI Figure S15), confirming photoluminescence loss post-fabrication, which can be caused by partial ligands removal after the stripping of the double-stack photoresist.

The cross-sectional SEM image in Figure 5g shows well-defined structures after the etching process, with an angle of approximately 80°. Confocal microscopy images show uniform green luminescence from both pristine and structured $FAPbBr_3$ NCs thin films (SI Figure S13e and Figure 5h, respectively). The blue shift in the PL emission after etching is further confirmed by the confocal fluorescence spectra (Figure S13d). Well-defined regular patterns of NC-based thin films with a thickness of approximately 230 nm (see SEM, optical microscope images and surface profile data in Figures S13f, g and h). A comparison of the two profilometer line scans before (Figure S13h) and after (Figure S13i) the nanofabrication process revealed a height difference of approximately 20 nm. We attribute this result to the bonding of carboxyl (COOH) groups in PMMA with the unpassivated ionic lead ($Pb^{2+}$) on the surface[79], resulting in the lifting of some surface layers of the NCs thin film during the PMMA removal. This could also account for the roughness reduction after the patterning process observed by AFM.

**Conclusions**

We developed a top-down patterning technique for Metal-halide perovskites based on conventional lithography and reactive ion etching. We demonstrated the method on organic-inorganic hybrids and all-inorganic compounds with different dimensionalities (3D, 2D, and 0D). The technique was evaluated via AFM, XRD, optical spectroscopy, and SEM, confirming that the phase and optical properties of the perovskites were not significantly affected by the patterning process, except at the edges of the patterned organic-inorganic metal halide perovskite. The organometal halide perovskite $MAPbI_3$ showed local halide exchange at the edges of the patterned area with the etching gas ions, which can be explained by its intrinsic soft lattice crystal structure. The developed microstructuring technique showed reproducible results with a resolution down to 1 µm, sufficient for industrial applications where perovskite could potentially be integrated in the future. The UV lithography combined with one step ICP-RIE developed in this work offers a favorable balance of resolution, material compatibility, scalability, and industrial feasibility. The presented approach is applicable on the wafer scale and can achieve high throughput and smaller dimensions with higher-resolution photolithography tools. Therefore, it can be used to monolithically integrate different types of metal-halide perovskites on chip with industry-standard processes. Our technology enables flexible device engineering and can accelerate the research and development of perovskite-based optoelectronics, electronics and energy harvesting applications.

## Methods

**Substrates:** Si/$SiO_2$ chips were diced and cleaned from the protective polymer layer by sonication in acetone and isopropanol, followed by 10 minutes of oxygen plasma cleaning in an RF plasma reactor. The chips were subsequently transferred into a nitrogen-filled glove box for the perovskite deposition processes.

**Perovskite Deposition:**

**All-inorganic perovskite thin films, namely,** CsBr (> 99.0%, TCI) and $PbBr_2$ (≥ 98%, Sigma Aldrich), were combined at a 1:1.7 molar ratio of 14% by weight, dissolved in dimethylsulfoxide (DMSO, anhydrous, ≥ 99.9%), and then left overnight at 60°C on a hot plate. The precursor solution was subsequently filtered via 0.20 µm pore size polytetrafluoroethylene (PTFE) filters. Finally, 80 µl of the precursor solution was spin-coated onto the cleaned Si/$SiO_2$ substrates at 2000 rpm for 40 seconds, followed by a thermal annealing process at 80°C for 10 minutes.

**Organic–inorganic hybrid perovskite thin films:** The $MAPbI_3$ perovskite solution was prepared with 1 M methylammonium ($CH_3NH_3I$, TCI), lead iodide ($PbI_2$, 99.999%, Sigma Aldrich), and dimethyl sulfoxide (DMSO, anhydrous, ≥ 99.9%). These compounds were dissolved in 1 ml of dimethylformamide (DMF, anhydrous, 99.8%) and heated to 60°C for 1 hour while stirring. The resulting solution was then filtered through a 0.20 µm pore size PTFE filter and spin-coated onto the cleaned Si/$SiO_2$ substrates at 6000 rpm for 30 seconds. Six seconds after the spin started, 300 µl of chlorobenzene (anhydrous, 99.8%) was added. The resulting film underwent a two-step annealing process (45°C for 5 minutes followed by 100°C for 10 minutes) to complete the conversion of the precursors into the $MAPbI_3$ perovskite layer.

**Quasi-2D perovskite thin films:** $PEA_2(MAPbBr_3)_{n-1}PbBr_4$ quasi-2D perovskite thin films were deposited via the spin coating technique. First, the perovskite solution was prepared by dissolving 0.4 mmol of methylammonium bromide (MABr, Sigma Aldrich), 0.6 mmol lead bromide ($PbBr_2$, TCI), 0.4 mmol phenylethylammonium bromide (PEABr, Sigma Aldrich), and 0.06 mmol methylammonium chloride (MACl, TCI) in 1 ml of DMSO. The prepared solution was stirred overnight at 60°C. 80 µl of filtered perovskite solution was spin-coated at 3000 rpm for 2 minutes and then annealed at 100°C for 30 minutes.

**Perovskite nanocrystalline thin films:** $FAPbBr_3$ NCs were synthesized via a hot-injection method at the Kovalenko laboratory via a previously published protocol[80]. The NC solution was dissolved in cyclohexane at a concentration of approximately 20 mg/ml. The solution was spin-coated at 2400 rpm for 30 seconds and then at 10000 rpm for 1 second.

**Photoluminescence Mapping:** Photoluminescence mapping was performed via a WITec confocal Raman microscope (model alpha300R). The samples were scanned with a 457 nm CW laser at 1 µW power. The measurements were taken at room temperature under ambient conditions.

**UV-Vis absorption spectroscopy:** The absorption spectrum of each sample before and after etching was recorded via ultraviolet-visible light spectrophotometry (PerkinElmer Lambda 1050) in the wavelength range of 300 to 800 nm.

**AFM, XRD, and SEM inspection:** AFM images were taken from each sample before and after etching via an atomic force microscope (Bruker Dimension Icon) in tapping mode. XRD with filtered Cu-Kα radiation (wavelength at 1.5405 Å) was carried out before and after the etching process via a PANalytical instrument at a current of 40 mA and a voltage of 40 kV. SEM micrographs were taken via an SEM Zeiss SUPRA 60 at 4 kV and a working distance of 3.5 mm.

**Confocal spectral fluorescence imaging:**

Confocal spectral fluorescence imaging was performed with an LSM 710 confocal microscope with a 34-channel Quasar detector (Zeiss, Oberkochen). The samples were excited with a 405 nm diode laser, and the emitted light was collected with an EC Plan-Neofluar 10x/0.30 objective. The microscope system was controlled, and images were acquired in λ mode with the ZEN black software version 2.3 SP1 FP3 (64-bit).

**Time-Correlated Single Photon Counting (TCSPC)**

The Time-Correlated Single Photon Counting (TCSPC) measurement was carried out with a self-constructed measurement system around an inverted microscope (TE300, Nikon). As a scanning system a confocal scan head (PMC2000, Nikon) was used. We have modified the excitation with a pulsed blue laser diode system and the detector with a single photon sensitive photomultiplier system. As excitation a pulsed laser (OEM module, PicoQuant) with a wavelength of 450 nm, a pulse duration below 200 ps, a repetition rate of 10 MHz and an average power of 1 mW was used. The emitted light from the sample was also recorded using the same confocal scan head and then detected using a photomultiplier (PMC 100, Becker & Hickl). For spectral filtering an optical bandpass filter (500-540 nm) was used in the scan head to separate the excitation light from the detection path. Time-correlated single photon counting was performed using a TCSPC measurement card (TimeHarp 200, PicoQuant). To check the homogeneity of the thin films, we used Fluorescence Lifetime Imaging Microscopy (FLIM). Typical image size is 128 by 128 $\mu m^2$ with the step size of 1 µm and a temporal resolution of better than 1 ns. The instrument response function (IRF) was measured using scattered laser light and has a full width at half maximum (FWHM) of 600 ps.

**Photoluminescence quantum yield (PLQY) measurements**

Absolute Photoluminescence: Excitation for the PL measurements was performed with a 445 nm CW laser (Insaneware) through an optical fibre into an integrating sphere. The intensity of the laser was adjusted to 20 $mW/cm^2$. A second optical fiber was used from the output of the integrating sphere to an Andor SR393i-B spectrometer equipped with a silicon CCD camera (DU420A-BR-DD, iDus). The system was calibrated by using a calibrated halogen lamp with specified spectral irradiance, which was shone into to integrating sphere. A spectral correction factor was established to match the spectral output of the detector to the calibrated spectral irradiance of the lamp. The spectral photon density was obtained from the corrected detector signal (spectral irradiance) by division by the photon energy (hf), and the photon numbers of the excitation and emission were obtained from numerical integration using Matlab[81].

**Acknowledgements**

This project has been funded by Deutsche Forschungsgemeinschaft within TRR 404 Active-3D (project number: 528378584), European Union's Horizon 2020 research and innovation programme under the Marie Skłodowska-Curie grant agreement No 956270, the project FOXES (951774), the German Research Foundation (DFG) through the project Hiper-Lase (GI 1145/4-1, LE 2440/12-1 and RI 1551/18-1) and RI 1551/22-1, and the German Ministry of Education and Research (BMBF) through the project NEPOMUQ (13N17112 and 13N17113). This work was also supported by the Confocal Microscopy Facility, a Core Facility of the Interdisciplinary Center for Clinical Research (IZKF) Aachen within the Faculty of Medicine at RWTH Aachen University. F.L. and S.Ö. thank the Volkswagen Foundation for Funding via the Freigeist Program. The authors thank P. Grewe and Dr. U. Böttger (Electronic Material Research Lab, RWTH Aachen University) for their support in the XRD measurements. The authors thank Dr. V. Morad and Prof. Dr. M. Kovalenko (ETH Zurich) for providing the perovskite nanocrystals.

**Supporting Information**

The Supporting Information is available free of charge on the ACS Publications website at DOI:

- Comparison of different photolithographic methods for metal halide perovskites (Table S1); grain size distribution histograms for $CsPbBr_3$ and $MAPbI_3$ thin films; large-area PL mapping, fluorescence and lifetime measurements, and PL quantum yield data for $CsPbBr_3$, $MAPbI_3$, quasi-2D $PEA_2(MAPbBr_3)_{n-1}PbBr_4$, and $FAPbBr_3$ nanocrystals before and after etching; SEM and optical images; PL spectrum comparisons between the centers and edges of etched structures; confocal microscopy images of chloride-, bromide-, and iodide-based perovskite compositions with varied features.

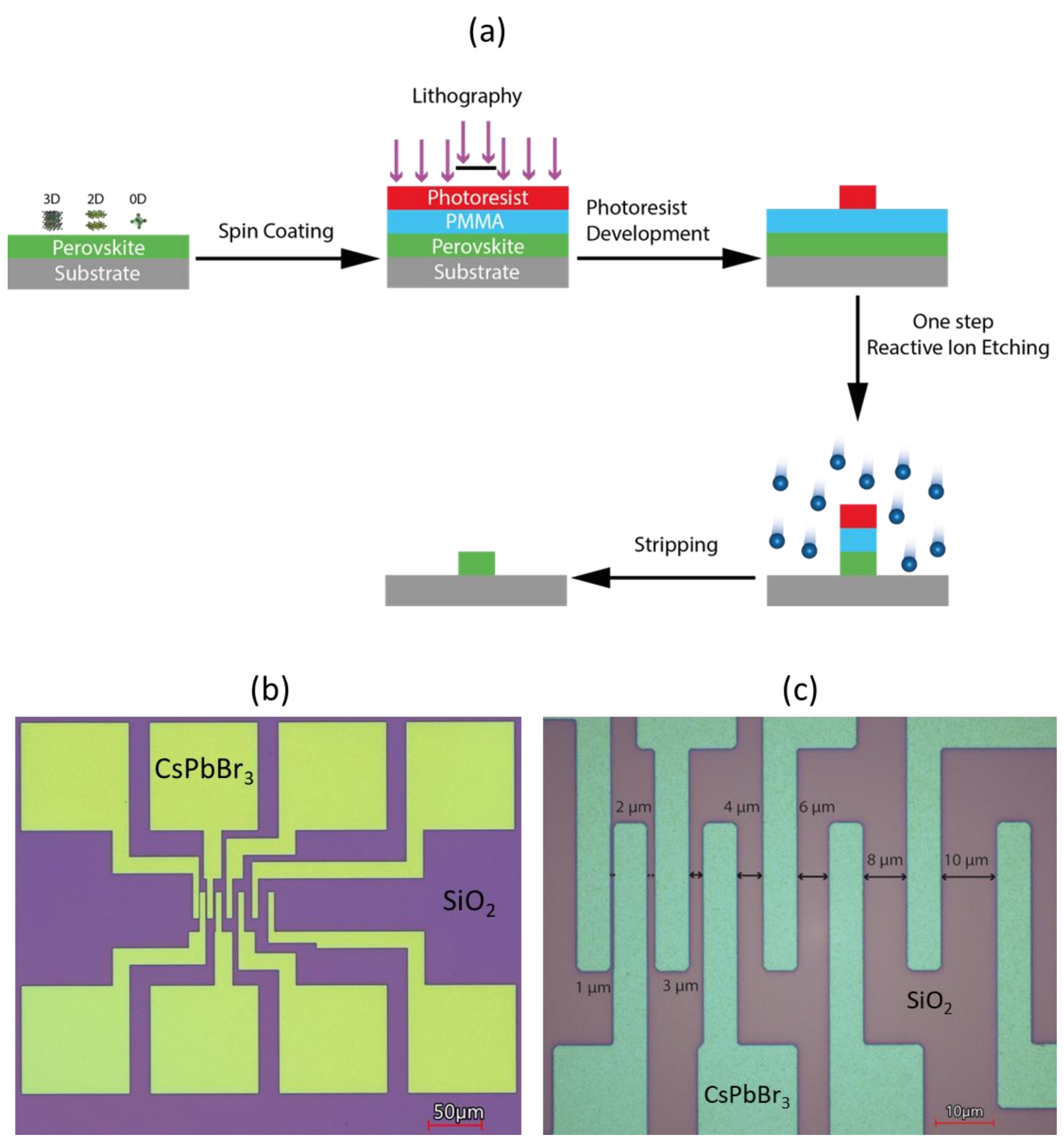

**Figure 1**. (a) The fabrication process includes the following steps: (1) Deposition of the perovskite thin film (3D, 2D or 0D), (2) Deposition of the PMMA layer and the photoactive resist layer (AZ MIR 701), (3) Development of the AZ MIR 701 resist layer (4) Pattern transfer into the perovskite layer by reactive ion etching (5) Stripping by dissolving the bottom resist layer and lifting the top layer. (b)-(c) Optical microscope image of device features achieved by top-down patterning method with a minimum feature dimension of 1 µm.

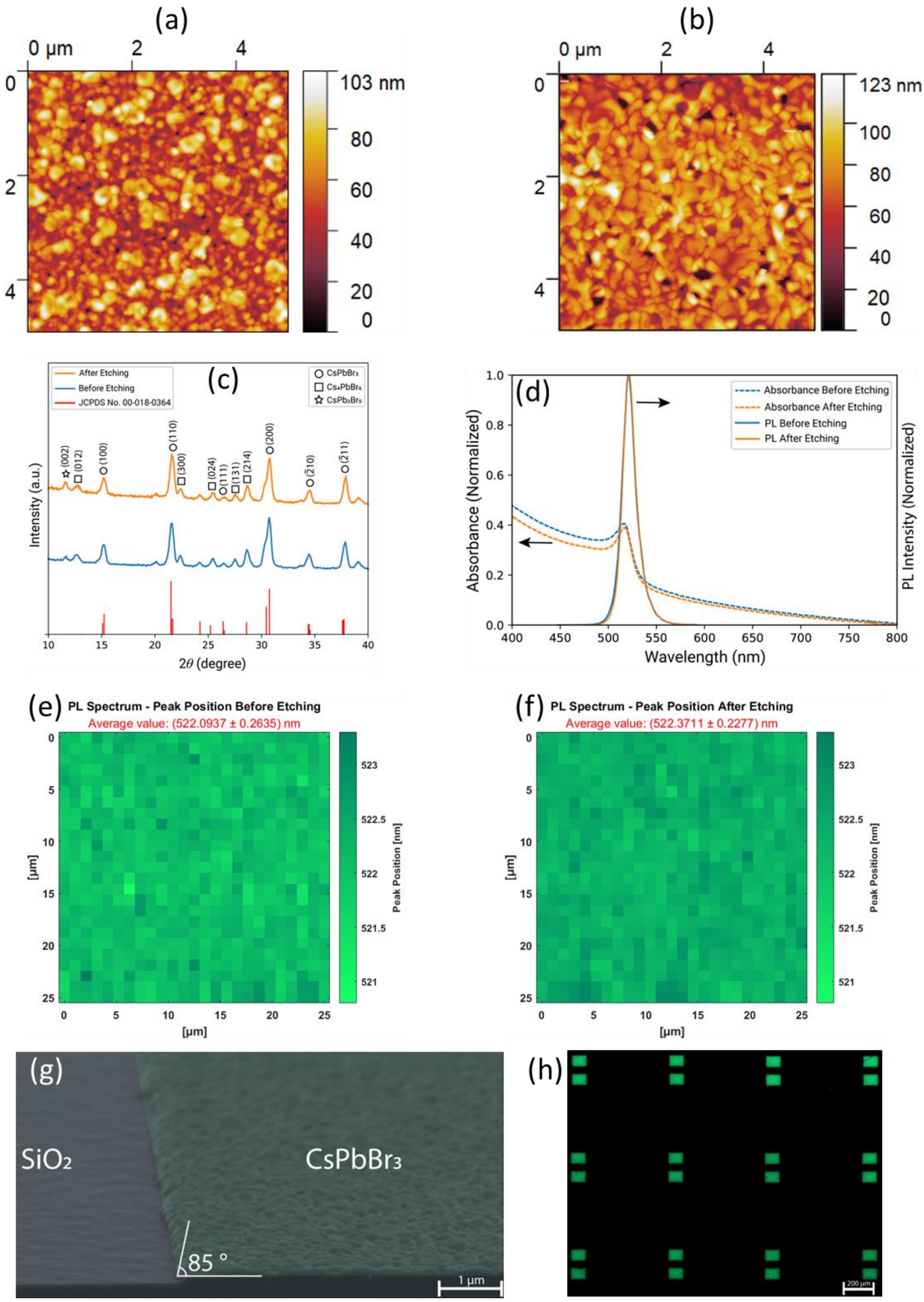

**Figure 2**. $CsPbBr_3$ patterning. (a) AFM scan of the pristine sample. (b) AFM scan after etching and stripping of the double stack. (c) X-ray diffraction patterns of the samples before and after etching. (d) UV–Vis spectra and PL spectra before and after etching. (e) – (f) Comparison of PL peak positions before and after etching over an area of 25 µm × 25 µm. (g) Cross-sectional SEM image of the etched structure. (h) Confocal fluorescence microscope image of an array of etched structures.

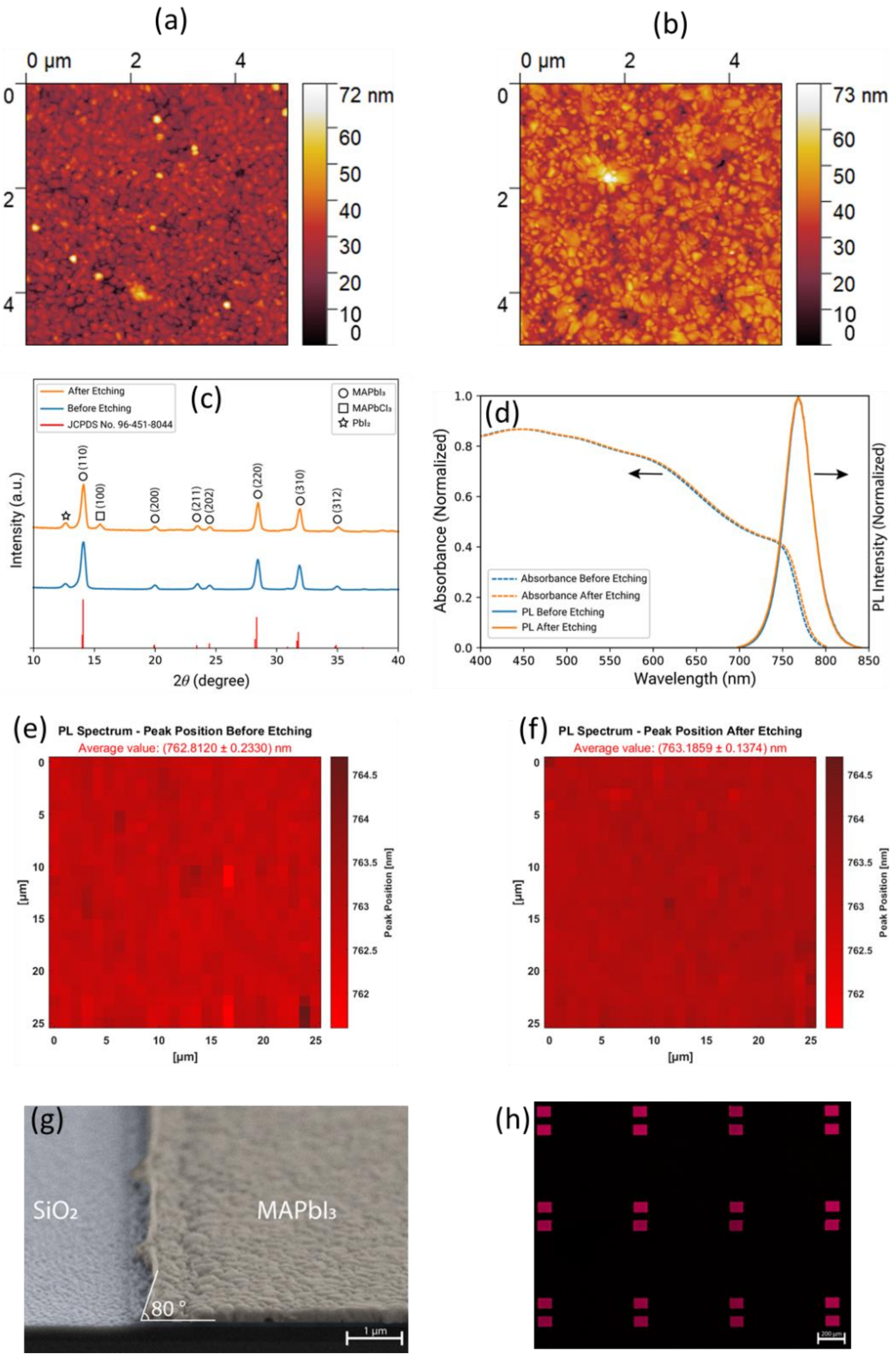

**Figure 3**. $MAPbI_3$ patterning. (a) AFM scan of the pristine sample. (b) AFM scan after etching and stripping of the double stack. (c) X-ray diffraction patterns of the samples before and after etching. (d) UV–Vis spectra and PL spectra before and after etching. (e) – (f) Comparison of PL peak positions before and after etching over an area of 25 µm × 25 µm. (g) Cross-sectional SEM image of the etched structure. (h) Confocal fluorescence microscope image of an array of etched structures.

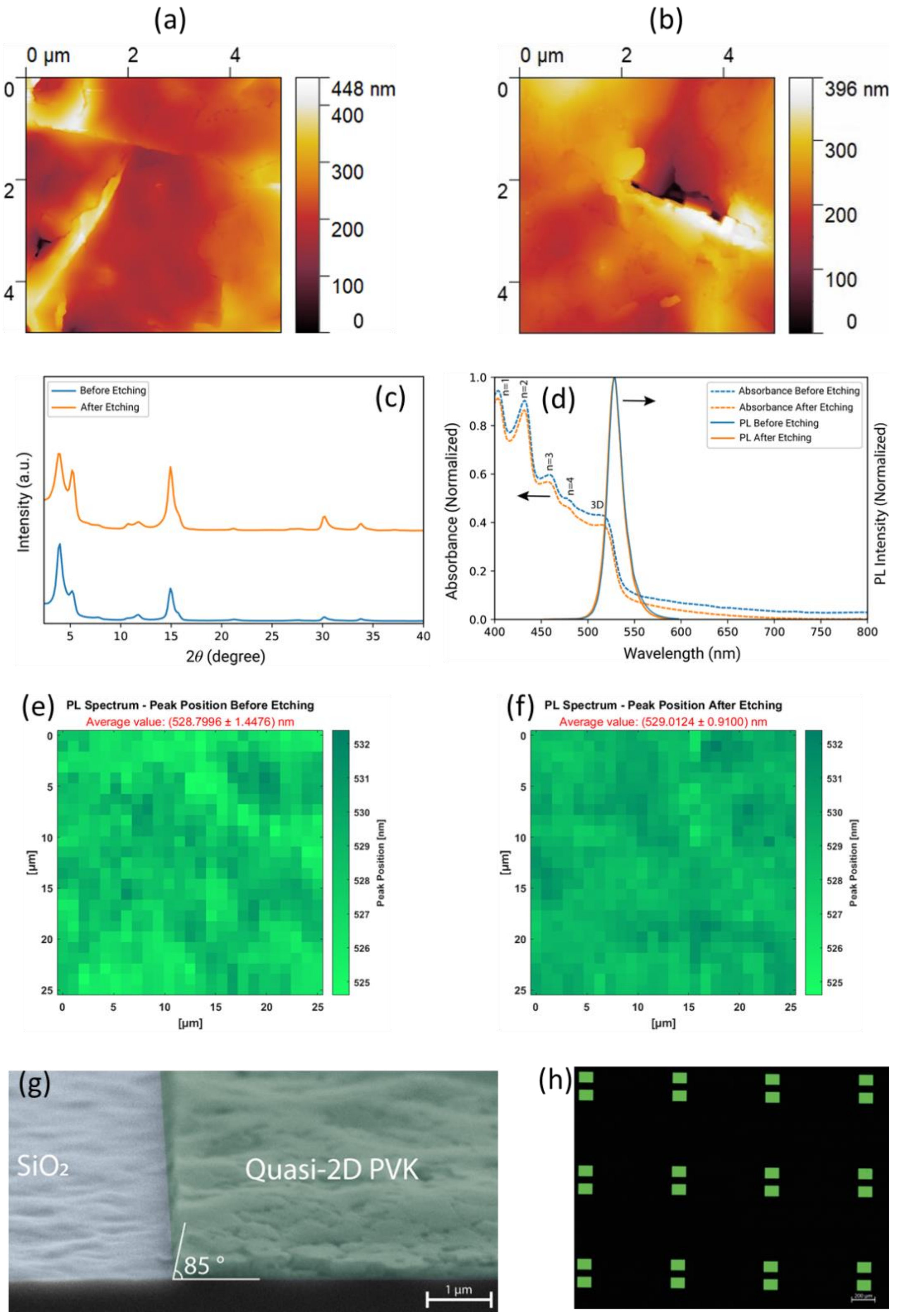

**Figure 4**. $PEA_2(MAPbBr_3)_{n-1}PbBr_4$ quasi-2D perovskite patterning. (a) AFM scan of the pristine sample. (b) AFM scan after etching and stripping of the double stack. (c) X-ray diffraction patterns of the samples before and after etching. (d) UV–Vis spectra and PL spectra before and after etching. (e) – (f) Comparison of PL peak positions before and after etching over an area of 25 µm × 25 µm. (g) Cross-sectional SEM image of the etched structure. (h) Confocal fluorescence microscope image of an array of etched structures.

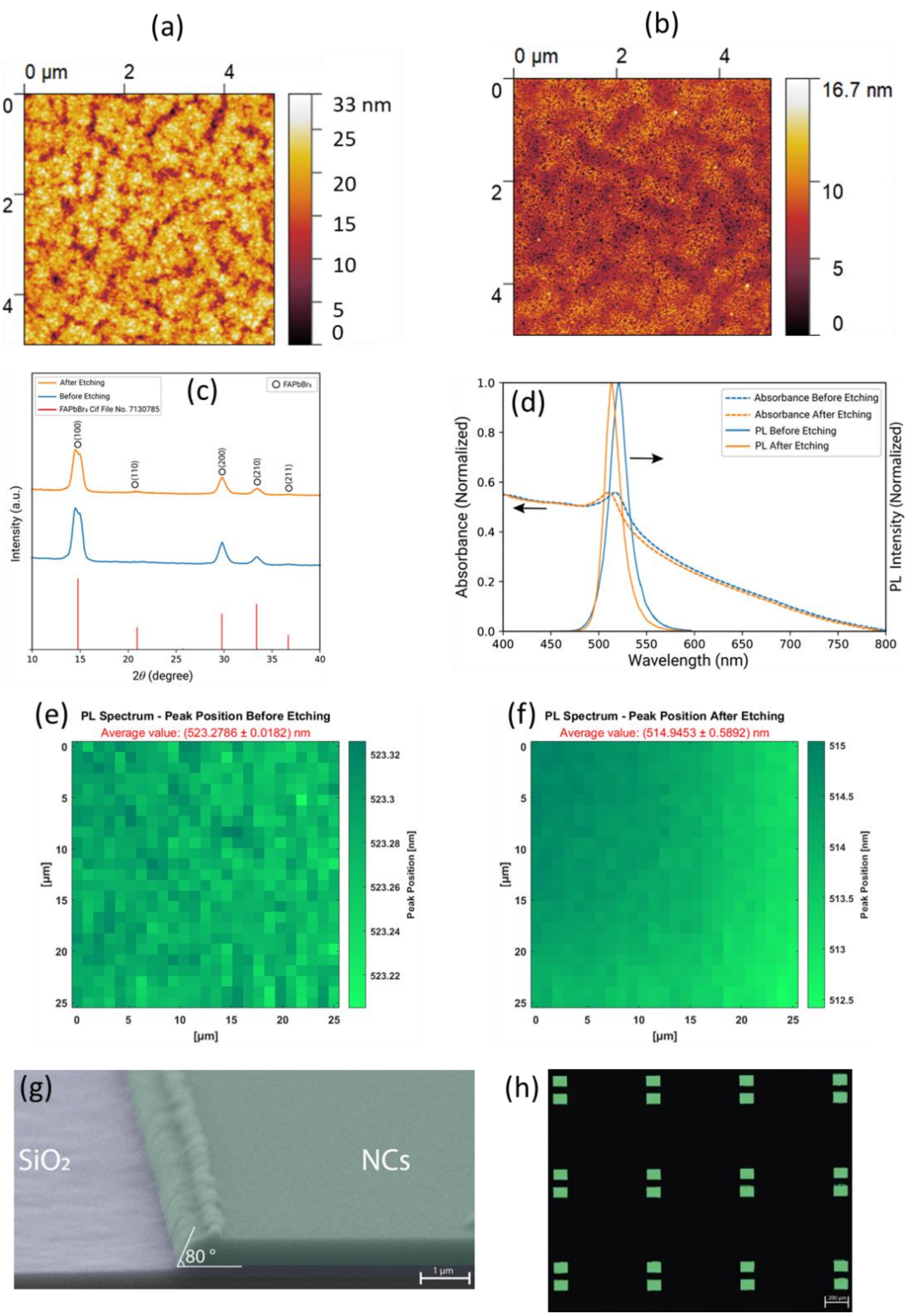

**Figure 5**. $FAPbBr_3$ patterning. (a) AFM scan map of the pristine sample. (b) AFM scan map after etching and stripping of the double stack. (c) X-ray diffraction patterns of the samples before and after etching. (d) UV–Vis spectra and PL spectra before and after etching. (e) – (f) Comparison of PL peak positions before and after etching over an area of 25 μm × 25 μm. (g) Cross-sectional SEM image of the etched structure. (h) Confocal fluorescence microscope image of an array of etched structures.

# Supporting information

## Versatile Top-Down Patterning Technique for Perovskite On-Chip Integration

*Federico Fabrizi*[1,2,˚], *Saeed Goudarzi*[2,˚], *Sana Khan*[1,2], *Tauheed Mohammad*[1], *Liudmila Starodubtceva*[2], *Piotr J. Cegielski*[1], *Felix Thiel*[3], *Sercan Özen*[4], *Maximilian Schiffer*[5,6], *Felix Lang*[4], *Peter Haring Bolívar*[3], *Thomas Riedl*[5,6], *Gerhard Müller-Newen*[7], *Surendra B. Anantharaman*[1,8], *Maryam Mohammadi*[1,*], *Max C. Lemme*[1,2,*]

[1] AMO GmbH, Otto-Blumenthal-Straße 25, Aachen, Germany.

[2] Chair of Electronic Devices, RWTH Aachen University, Otto-Blumenthal-Straße 25, Aachen, Germany.

[3] Institute for High Frequency and Quantum Electronics, University of Siegen, 57076 Siegen, Germany.

[4] Institute Freigeist Juniorgroup, Radiation Tolerant Electronics with Soft Semiconductors (ROSI), University of Potsdam, Karl-Liebknecht-Straße 24/25, Potsdam-Golm, Germany.

[5] Institute of Electronic Devices, University of Wuppertal, Rainer-Gruenter-Str. 21, 42119 Wuppertal, Germany.

[6]Wuppertal Center for Smart Materials & Systems (CM@S), University of Wuppertal, *Rainer-Gruenter-Str. 21,* 42119 Wuppertal, Germany

[7]Institute of Biochemistry and Molecular Biology, Uniklinik RWTH Aachen, Pauwelsstrasse 30, Aachen, Germany.

[8]Low-dimensional Semiconductors Lab, Department of Metallurgical and Materials Engineering, Indian Institute of Technology Madras, Chennai 600036, India.

° These authors contribute equally, fabrizi@amo.de, saeed.goudarzi@eld.rwth-aachen.de

‖ Corresponding author, mohammadi@amo.de, max.lemme@eld.rwth-aachen.de

**Table S1**. Comparison of the state-of-the-art photolithographic methods used with metal halide perovskite.

| Patterning Method | Resolution Limit | Scalability | Material Compatibility | Industrial Feasibility |
|---|---|---|---|---|
| **Contact Photolithography + ICP-RIE (this work)** | **~1 µm** | **High** | **Excellent** | **High** |
| Electron beam lithography (EBL)[1] | <10 nm | Low | Good | Expensive, Low Throughput |
| Nanoimprint Lithography (NIL)[2,3] | ~20 nm | Medium | Moderate | Moderate |
| Focused Ion Beam (FIB)[4] | <10 nm | Low | Moderate | Low Throughput |
| Laser writing[5,6] | <200 nm | Medium | Moderate | Moderate |

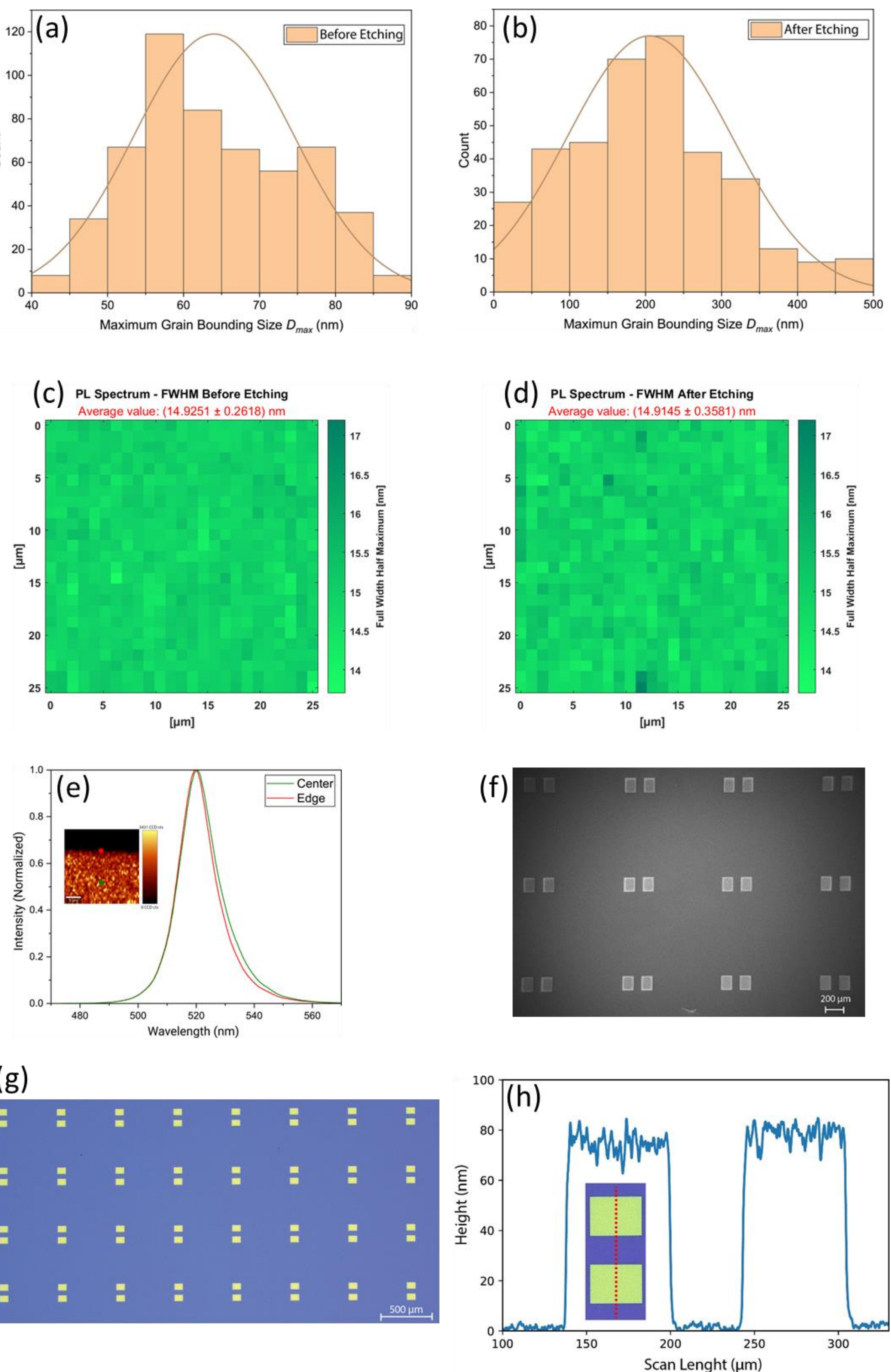

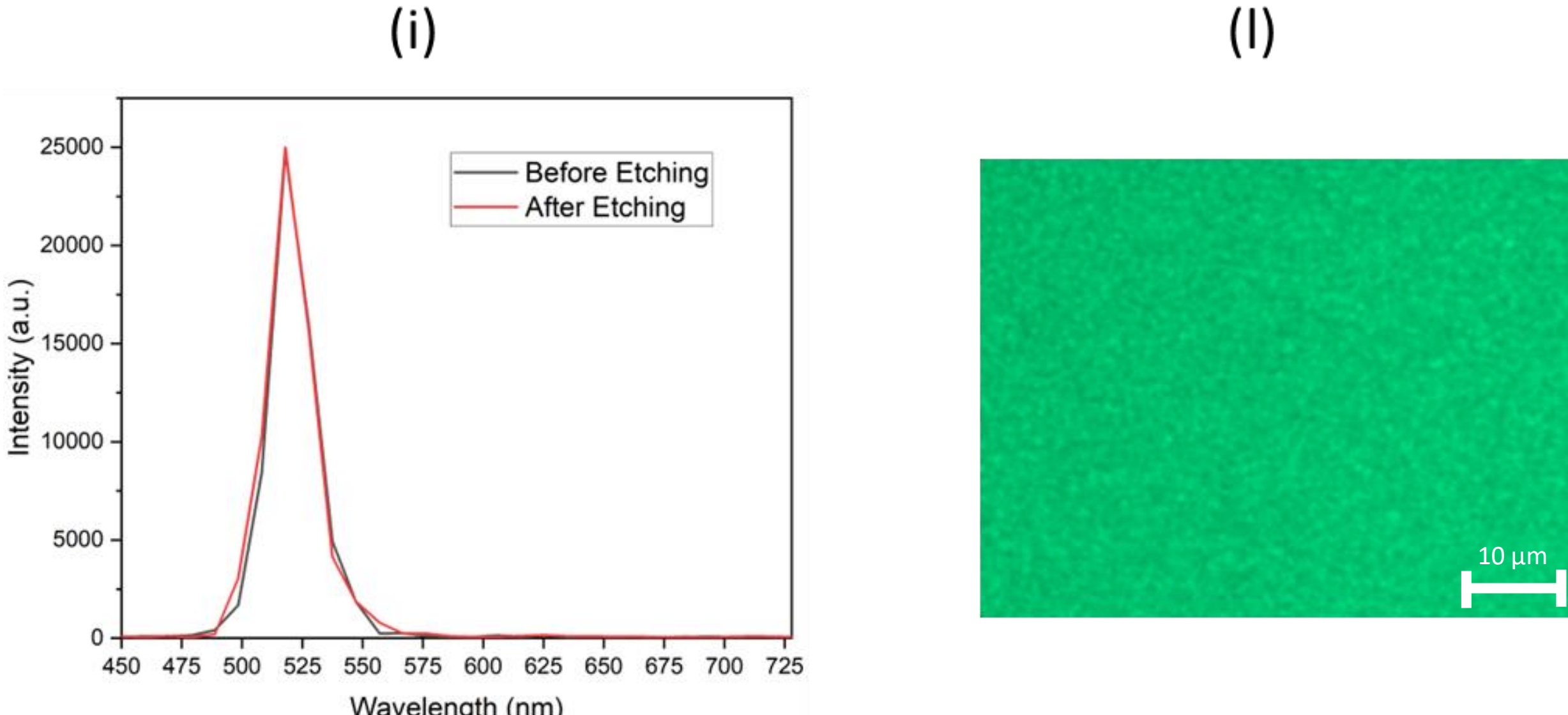

**Figure S1**. $CsPbBr_3$ patterning supplementary information. (a)-(b) Histogram plots of the grain size distribution before and after the etching process. (c)-(d) Large area scans of 25 µm x 25 µm before and after etching, where each pixel corresponds to a fitted value of the FWHM of the PL spectrum. (e) PL spectrum comparison between the edge and center of the etched structure. (f) Top-view SEM image. (g) Optical microscope image of the patterned structures. (h) Surface profilometer line scan of the etched structures after the stripping of the photoresist. (i) Fluorescence spectra before and after the etching. (l) Confocal image of the pristine $CsPbBr_3$ sample.

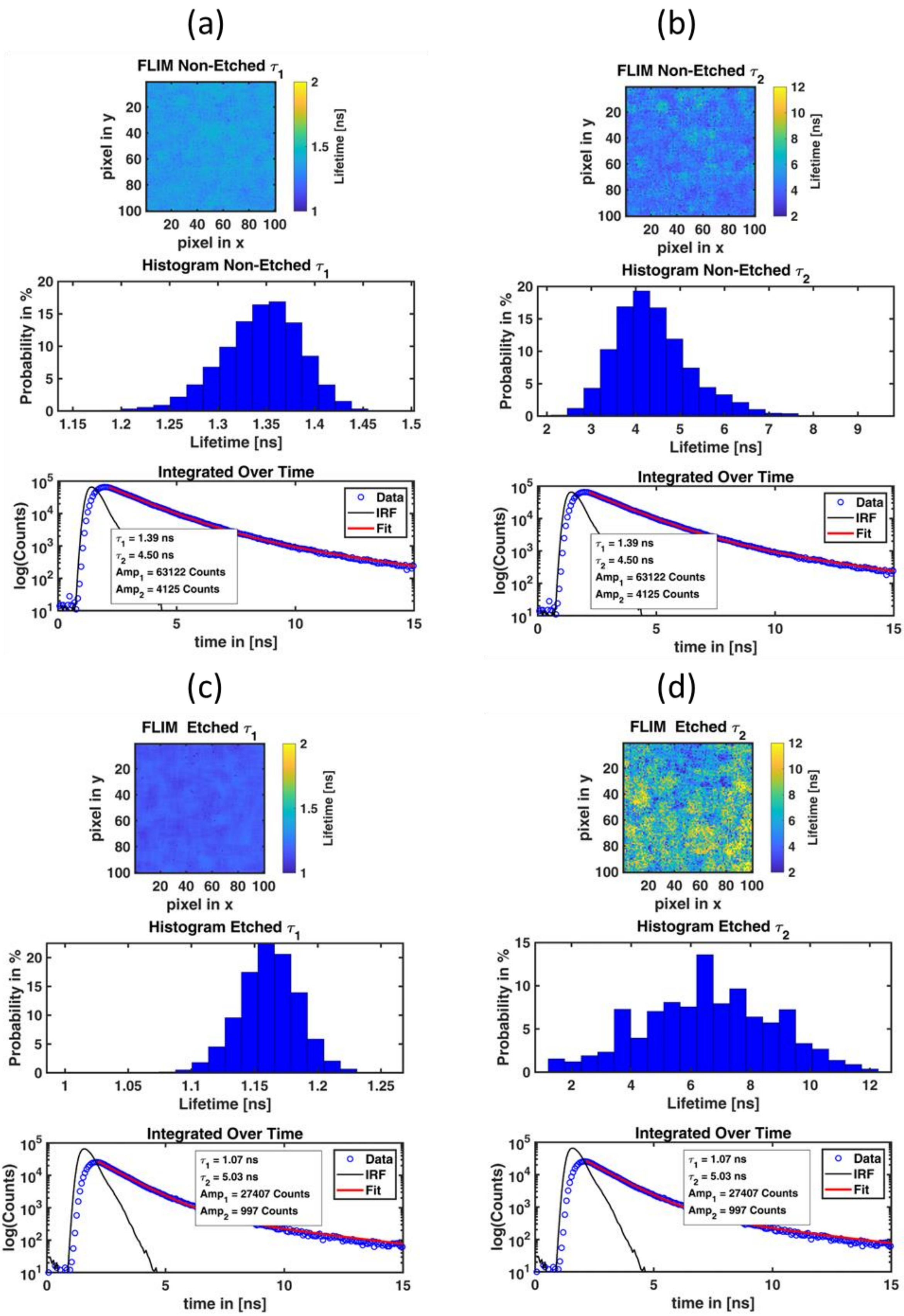

**Figure S2**. $CsPbBr_3$ lifetime measurements. (a)-(b) Photoluminescence (PL) decay time $\tau_1$ and $\tau_2$ for the pristine sample. (c)-(d) PL decay time $\tau_1$ and $\tau_2$ for the etched sample.

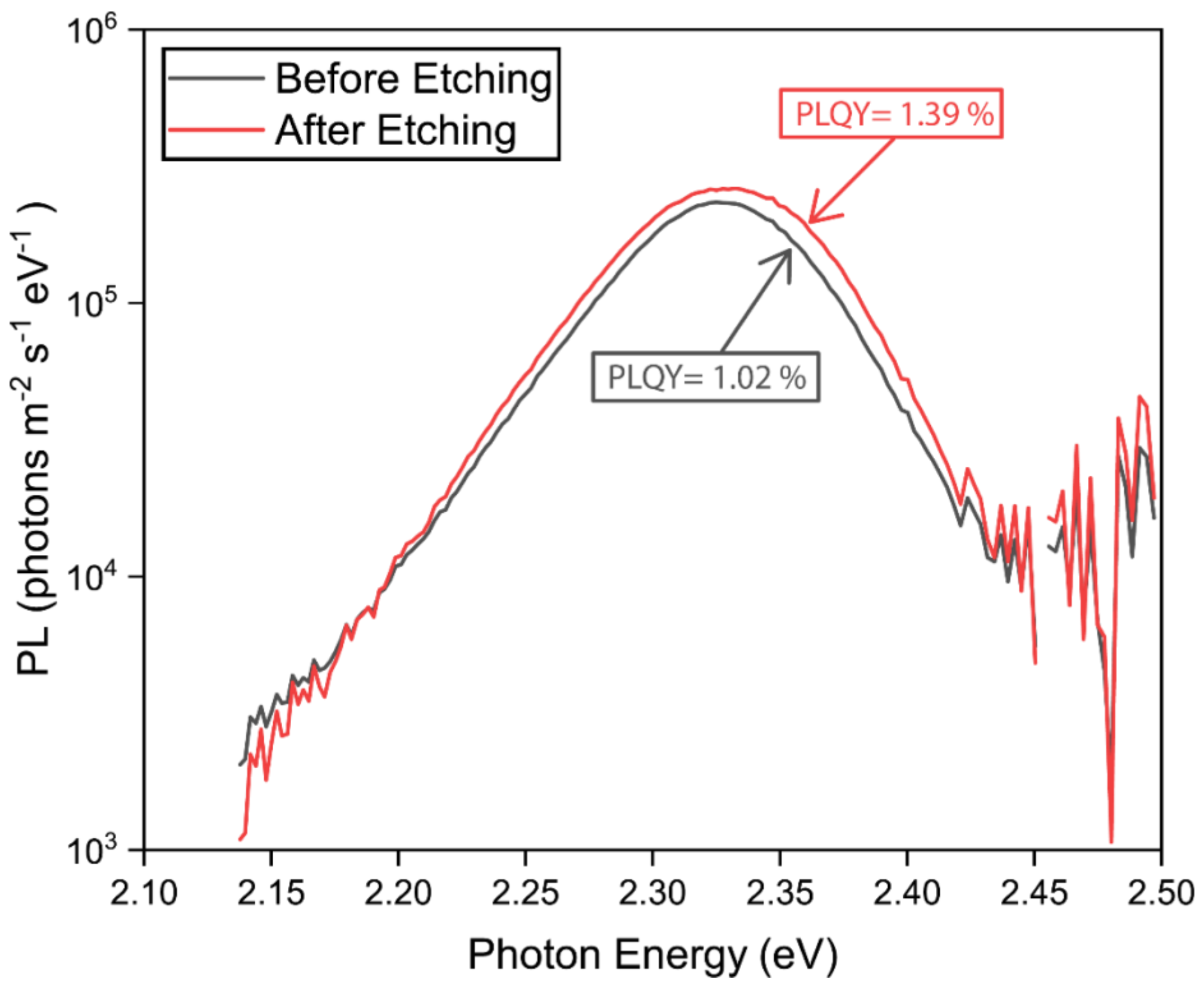

**Figure S3**. $CsPbBr_3$ PLQY measurements before and after the etching process.

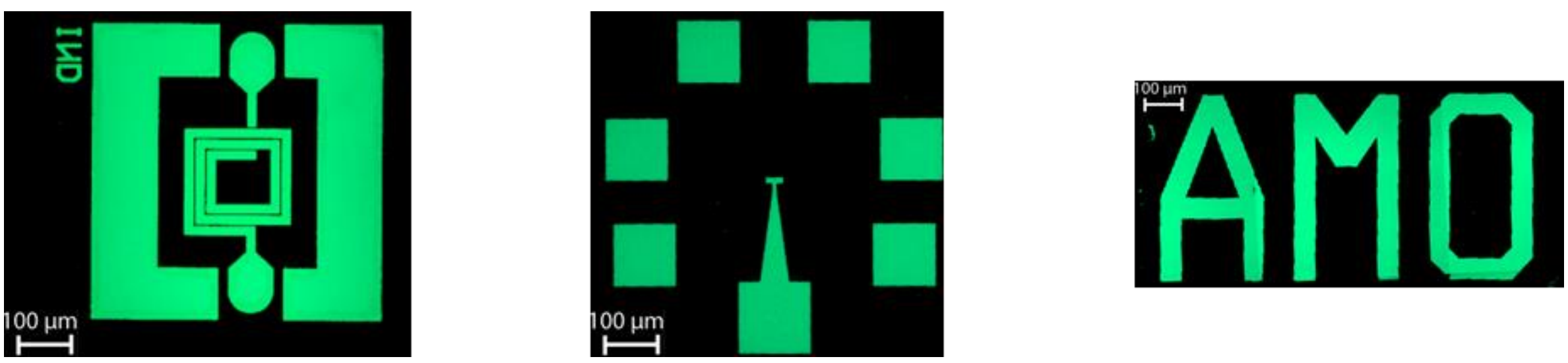

**Figure S4**. $CsPbBr_3$ confocal microscopy images of different features.

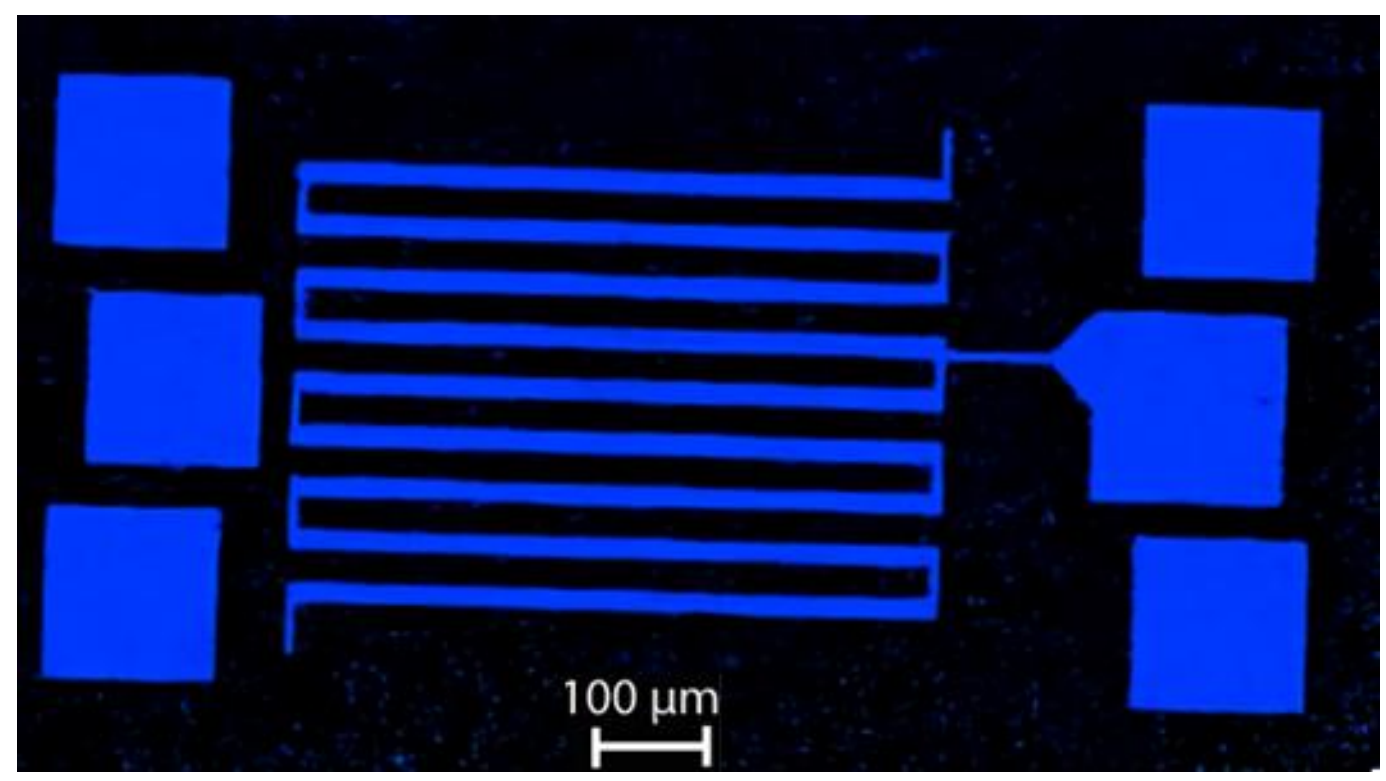

**Figure S5**. $CsPbCl_3$ confocal microscopy image.

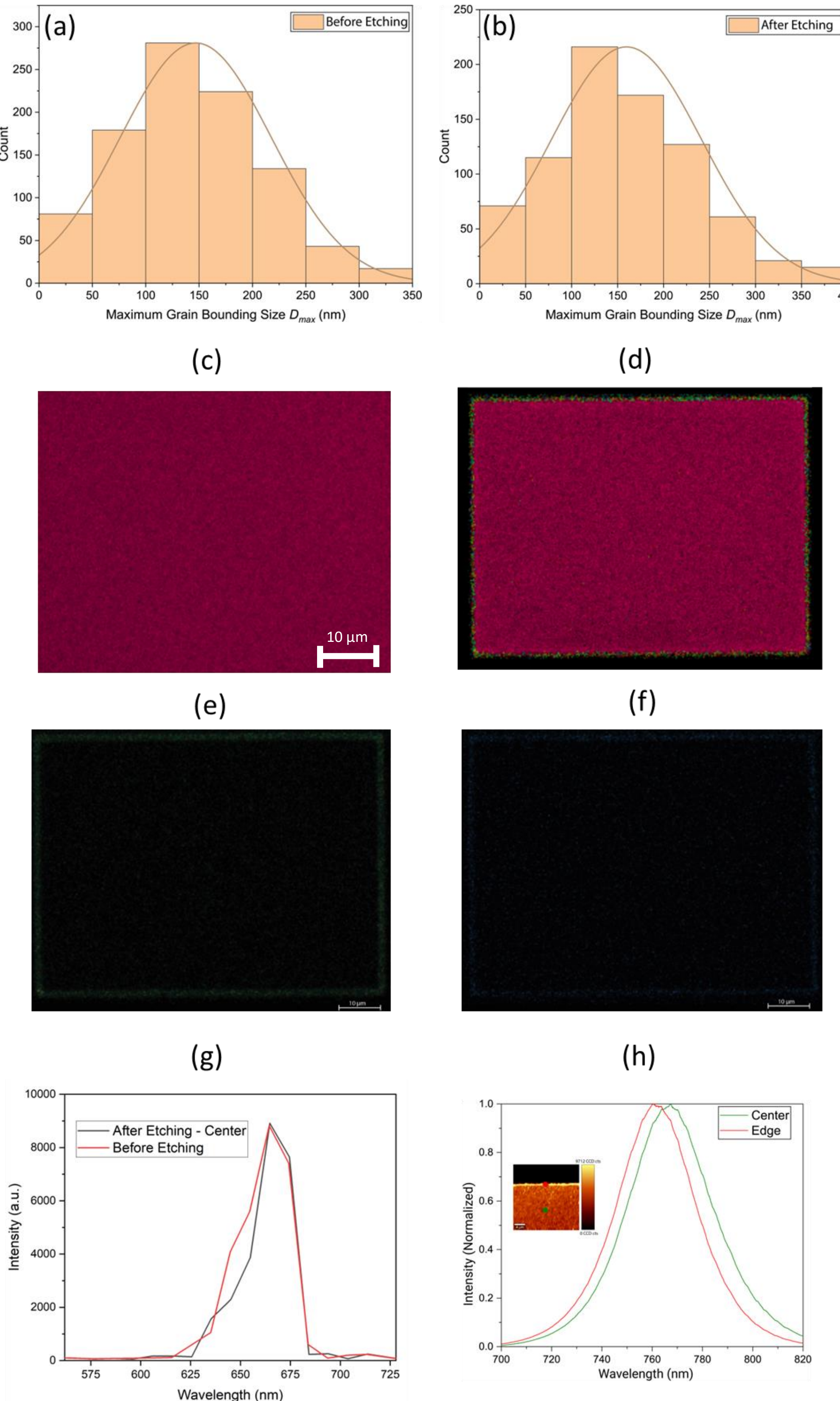

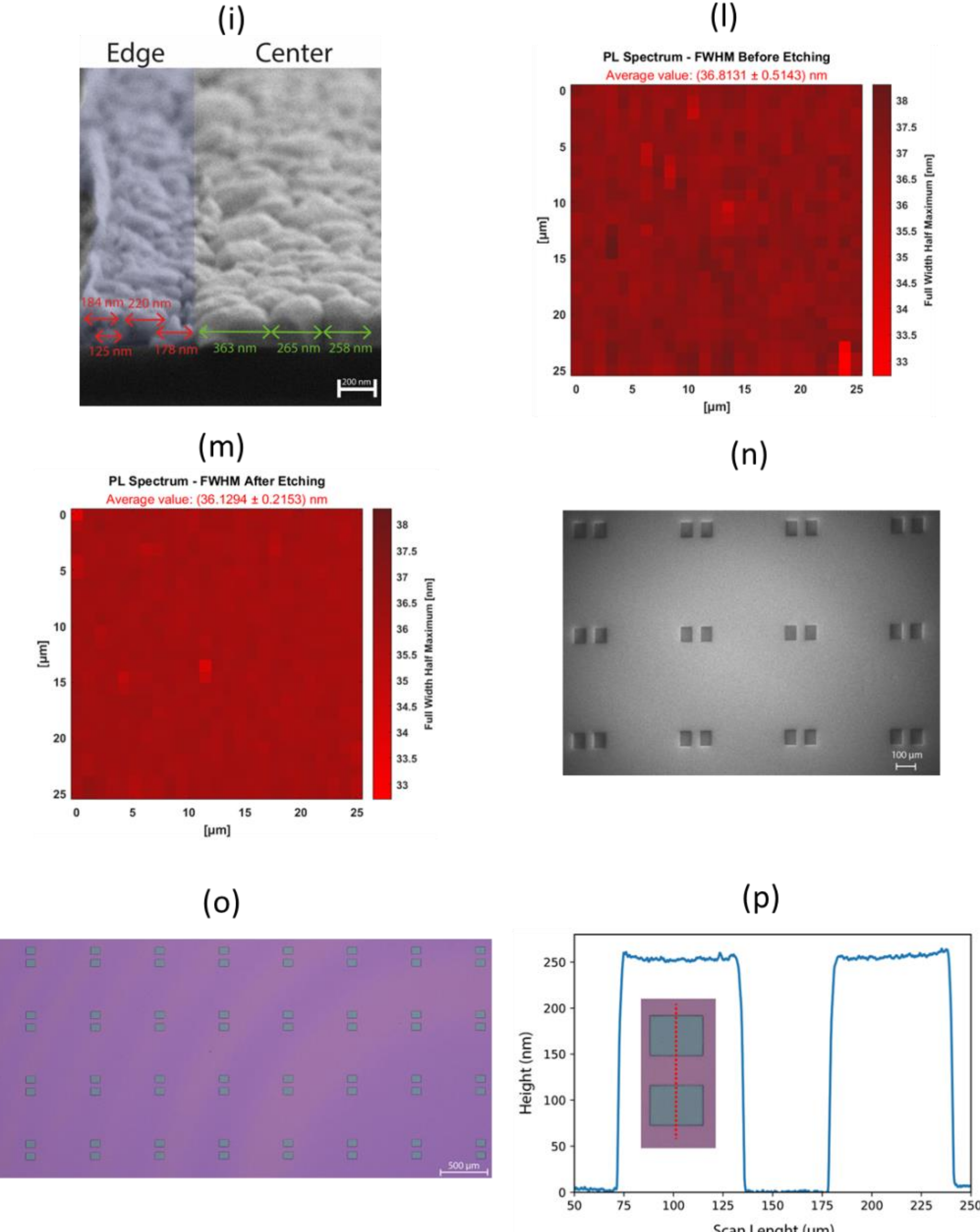

**Figure S6**. $MAPbI_3$ patterning supplementary information. (a)-(b) Histogram plots of the grain size distribution before and after the etching process. (c) Confocal image of the pristine $MAPbI_3$ sample. (d) Confocal image of one etched structure. A blue/green shallowness around the edges is visible. This is a confirmation of the local doping of the $MAPbI_3$ due to the etching gases used during the process. (e)-(f) Green and blue emission components coming from a mixed halide phase forming during the process. (g) Fluorescence spectra of the spin-coated $MAPbI_3$ and after the etching process. (h) PL spectrum comparison between the edge and center of the etched structure. A clear blue shift in the PL peak position on the edge is visible. (i) SEM cross-section image, where is visible a clear reduction in the grain sizes between the edge and the center of the etched area. (l)-(m) Large area scans of 25 µm x 25 µm before and after etching. Each pixel corresponds to a fitted value of the FWHM of the PL spectrum. (n) Top-view SEM image. (o) Optical microscope image of the patterned structures. (p) Surface profilometer line scan of the etched structures after the stripping of the photoresist.

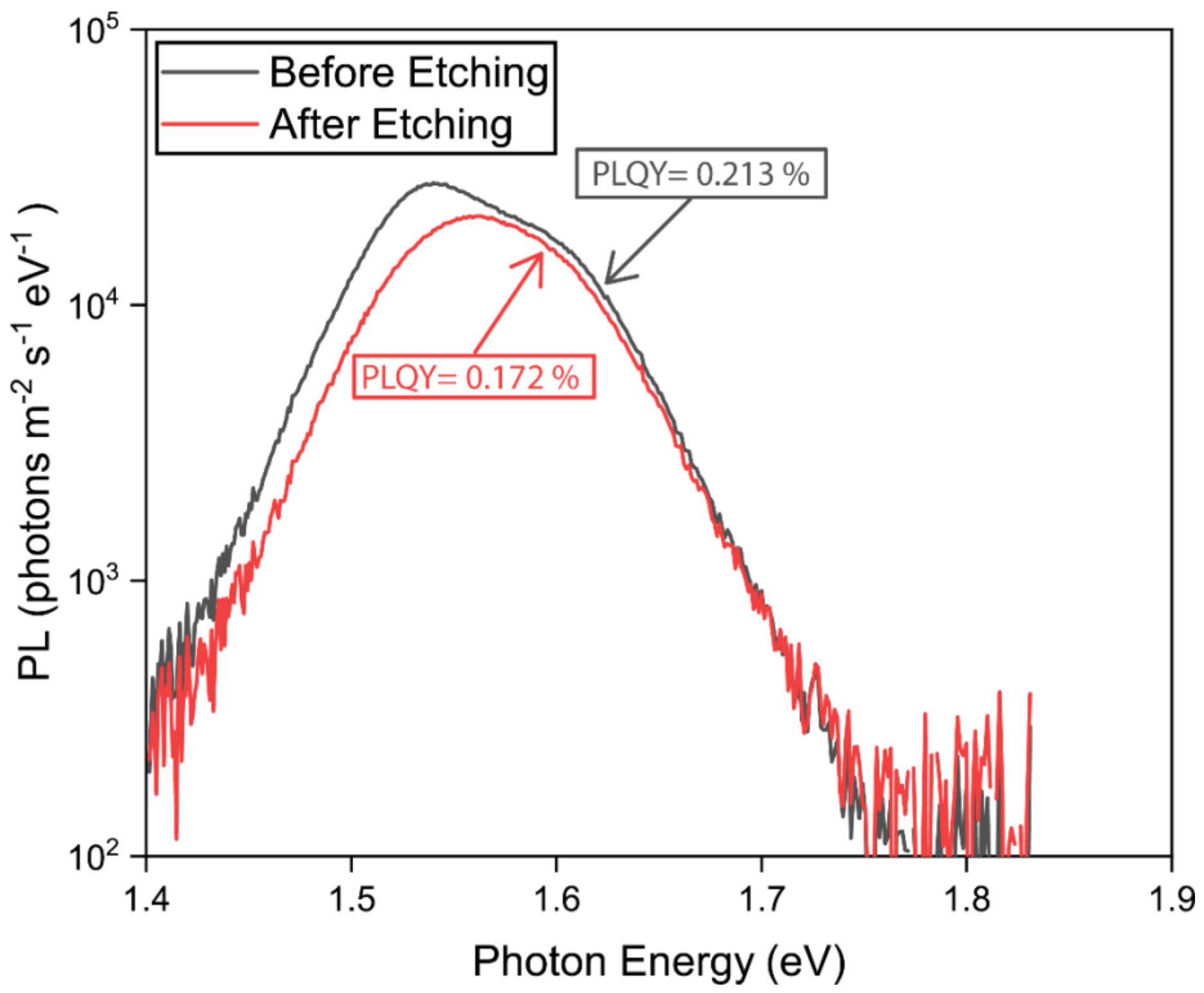

**Figure S7**. $MAPbI_3$ PLQY measurements before and after the etching process.

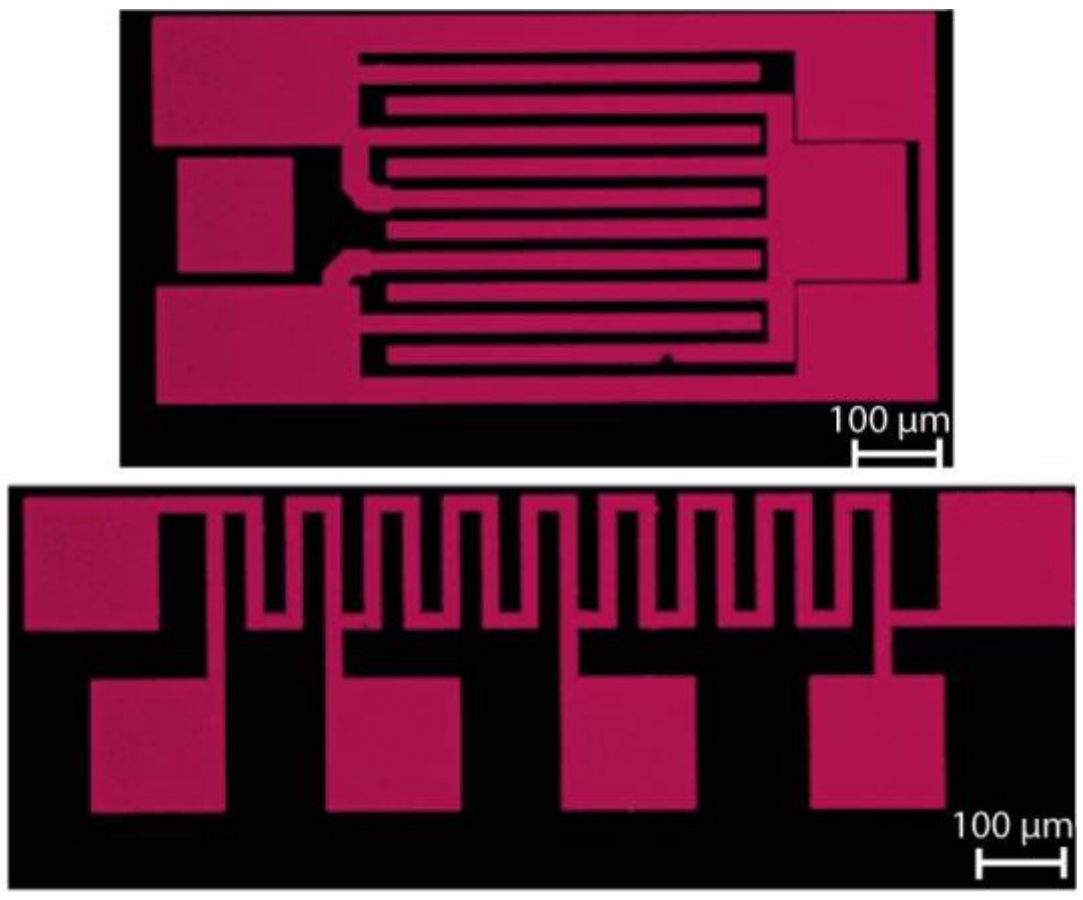

**Figure S8**. $MAPbI_3$ confocal microscopy images of different shapes and patterns.

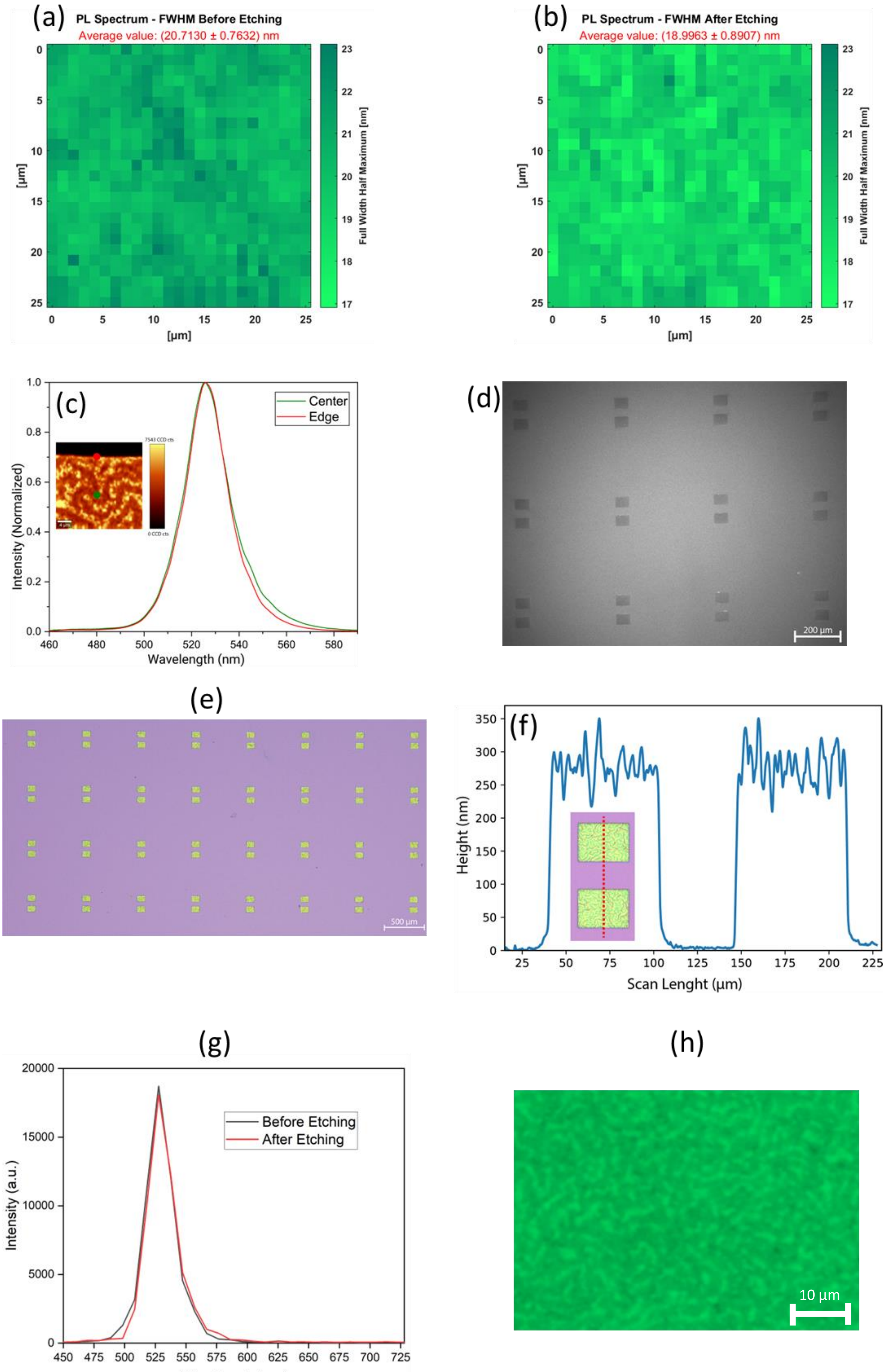

**Figure S9**. $PEA_2(MAPbBr_3)_{n-1}PbBr_4$ quasi-2D patterning supplementary information. (a)-(b) Large area scans of 25 µm x 25 µm before and after etching, where each pixel corresponds to a fitted value of the FWHM of the PL spectrum. (c) PL spectrum comparison between the edge and center of the etched structure. (d) Top-view SEM image. (e) Optical microscope image of the patterned structures. (f) Surface profilometer scan line of the etched structures after the stripping of the photoresist. (g) Fluorescence spectra before and after the etching. (h) Confocal image of the pristine $PEA_2(MAPbBr_3)_{n-1}PbBr_4$ sample.

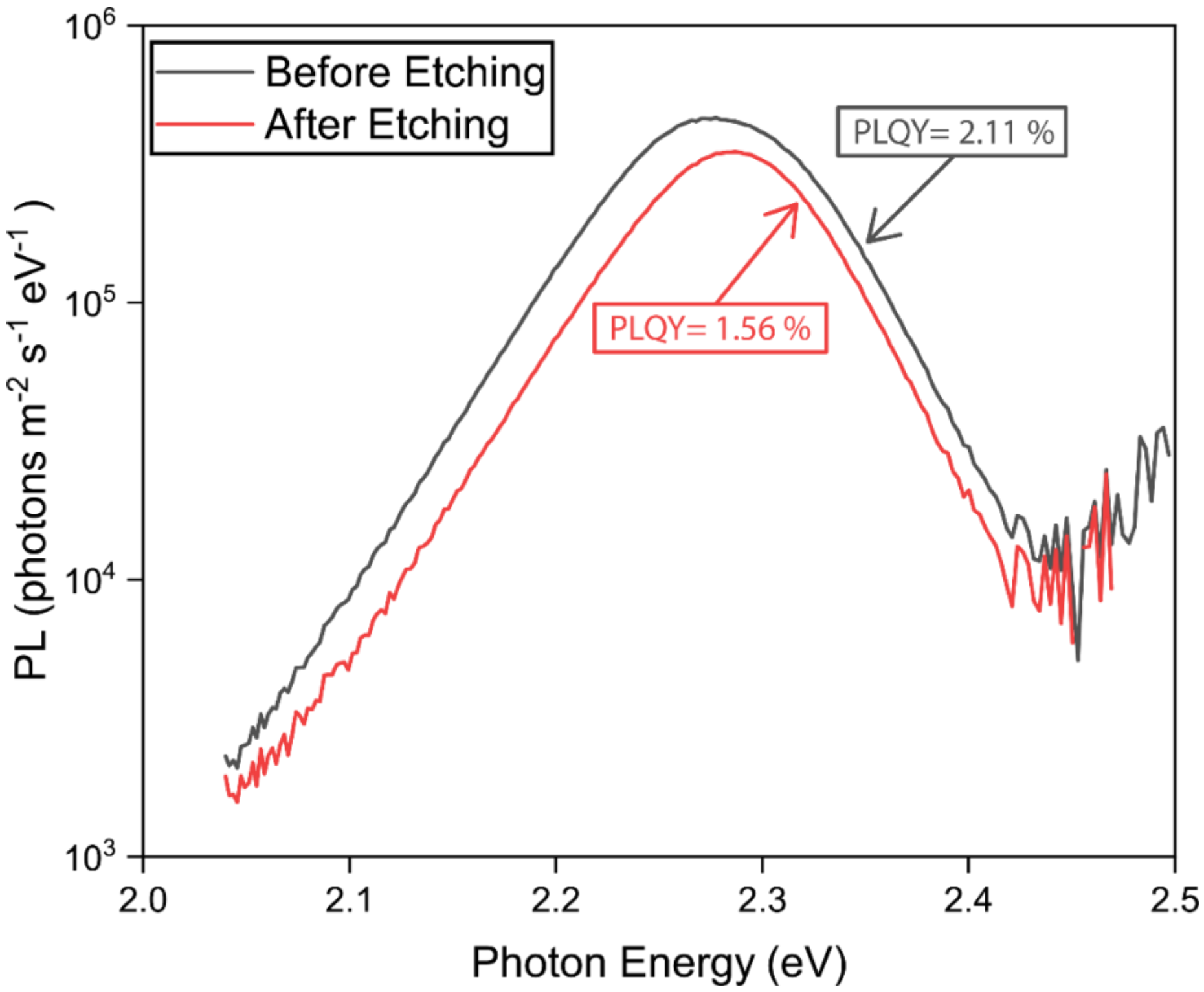

**Figure S10**. $PEA_2(MAPbBr_3)_{n-1}PbBr_4$ PLQY measurements before and after the etching process.

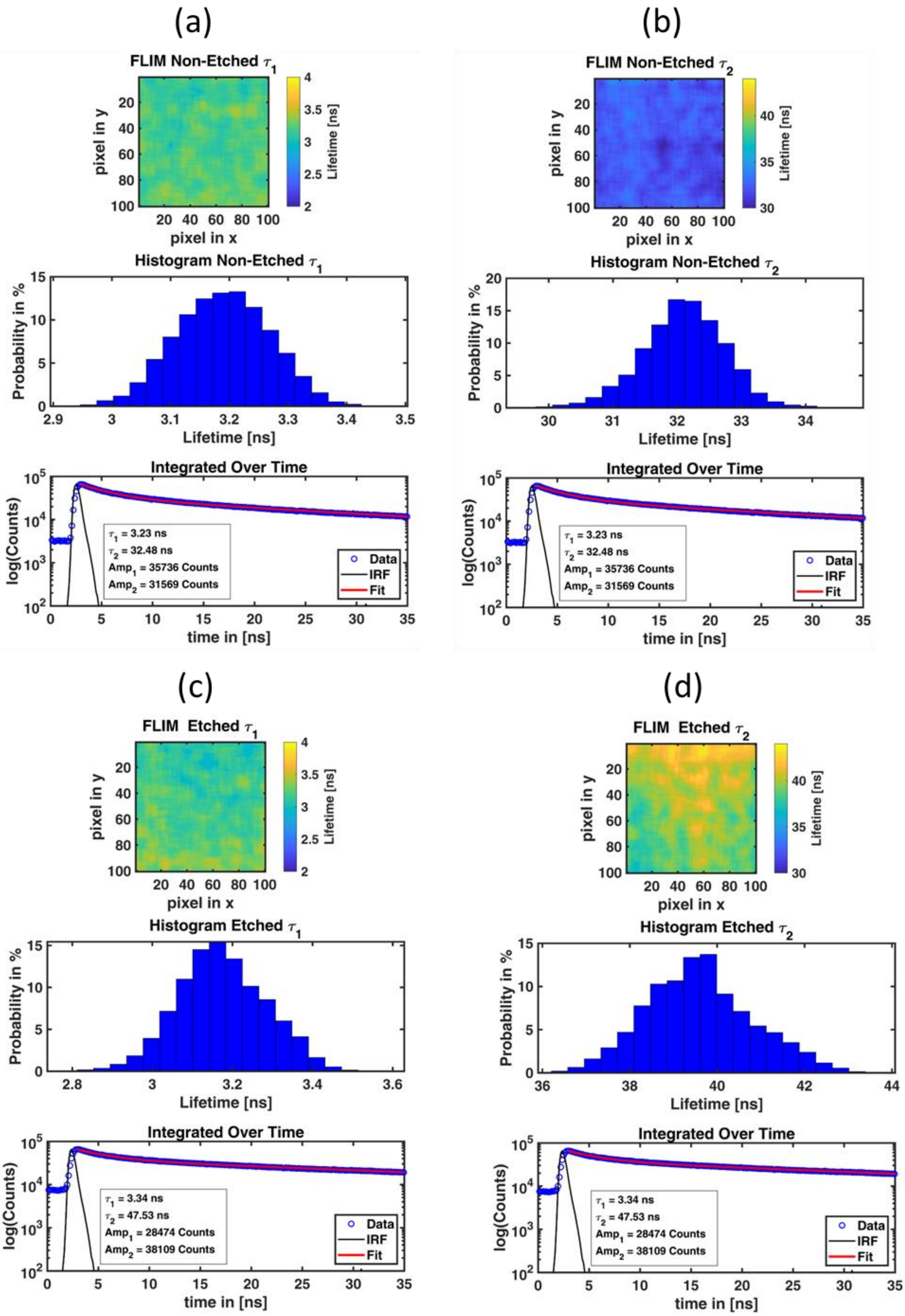

**Figure S11**. $PEA_2(MAPbBr_3)_{n-1}PbBr_4$ lifetime measurements. (a)-(b) Photoluminescence (PL) decay time $\tau_1$ and $\tau_2$ for the pristine sample. (c)-(d) PL decay time $\tau_1$ and $\tau_2$ for the etched sample.

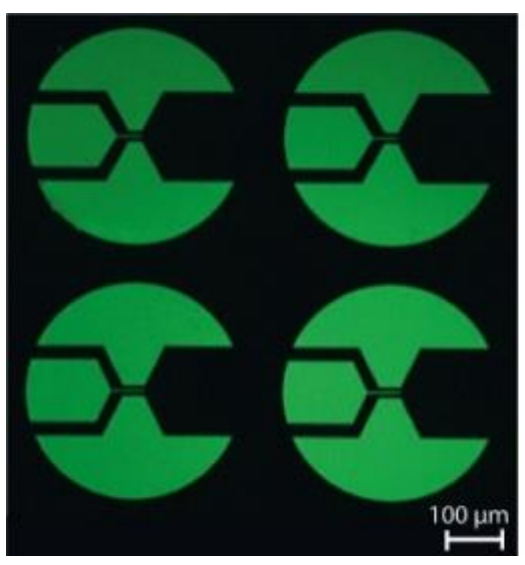

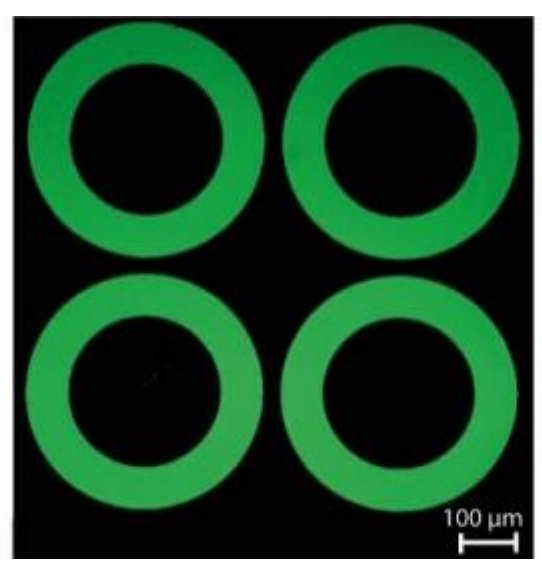

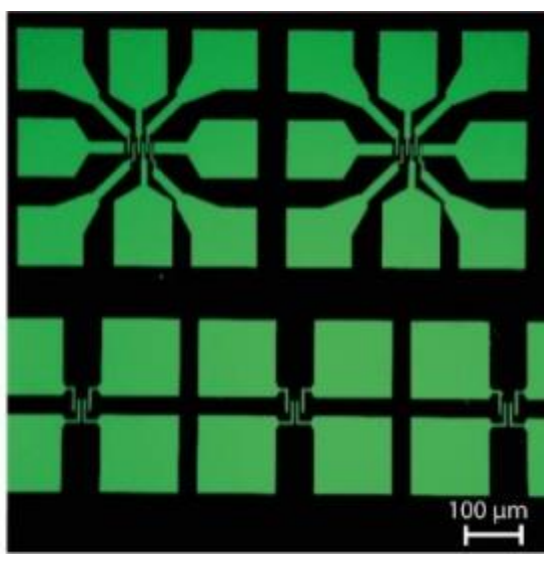

**Figure S12**. $PEA_2(MAPbBr_3)_{n-1}PbBr_4$ confocal microscopy images of different shapes and patterns.

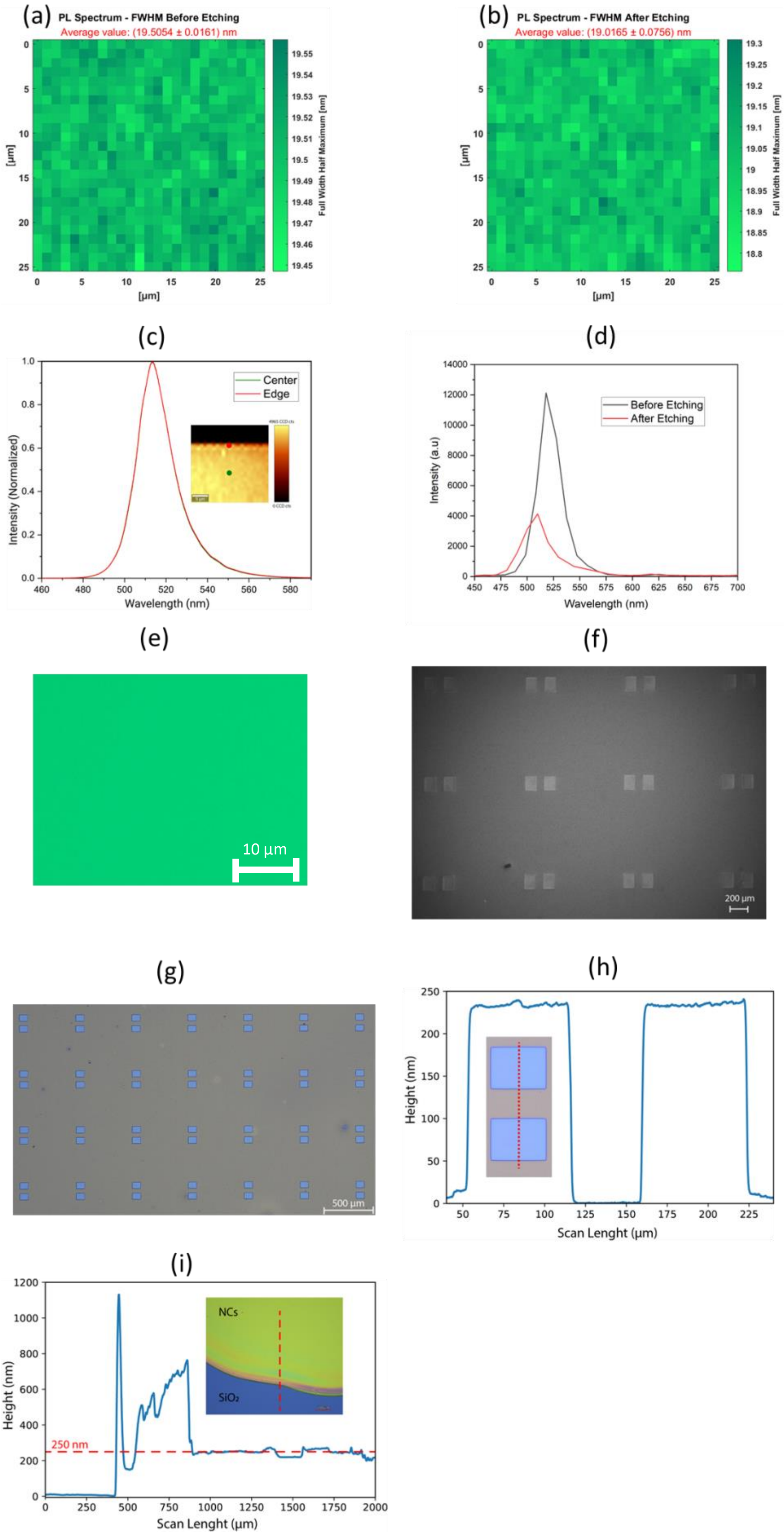

**Figure S13**. $FAPbBr_3$ NCs patterning supplementary information. (a)-(b) Large area scans of 25 μm x 25 μm before and after etching, where each pixel corresponds to a fitted value of the FWHM of the PL spectrum. (c) PL spectrum comparison between the edge and center of the etched structure. (d) Fluorescence spectra before and after the etching. (e) Confocal image of the pristine $FAPbBr_3$ NCs sample. (f) Top-view SEM image. (g) Optical microscope image of the patterned structures. (h) Surface profilometer scan line of the etched structures after the stripping of the photoresist. (i) Surface profilometer scan line of the etched structures after the spin coating and before the nanofabrication process.

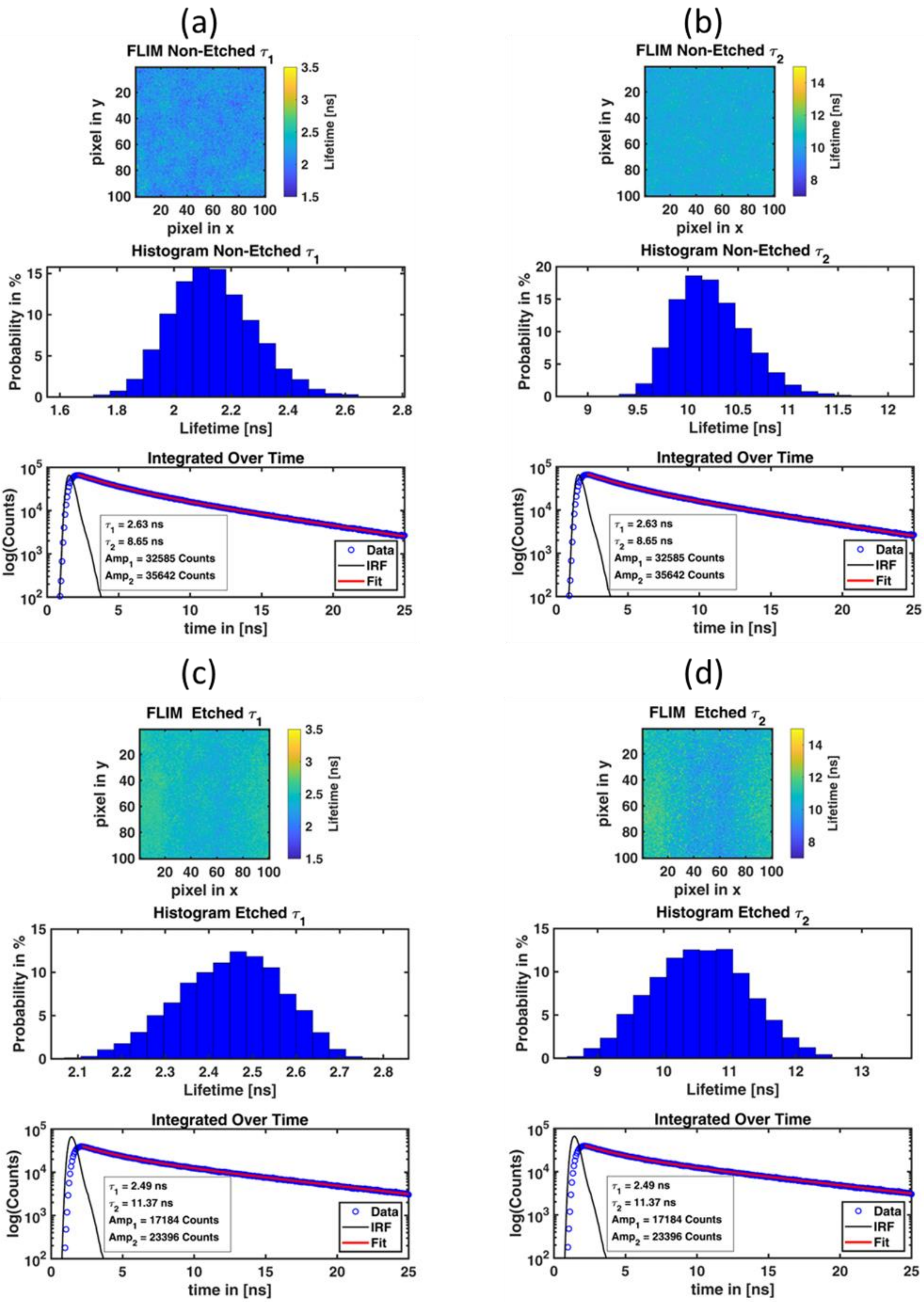

**Figure S14**. $FAPbBr_3$ NCs lifetime measurements. (a)-(b) Photoluminescence (PL) decay time $\tau_1$ and $\tau_2$ for the pristine sample. (c)-(d) PL decay time $\tau_1$ and $\tau_2$ for the etched sample.

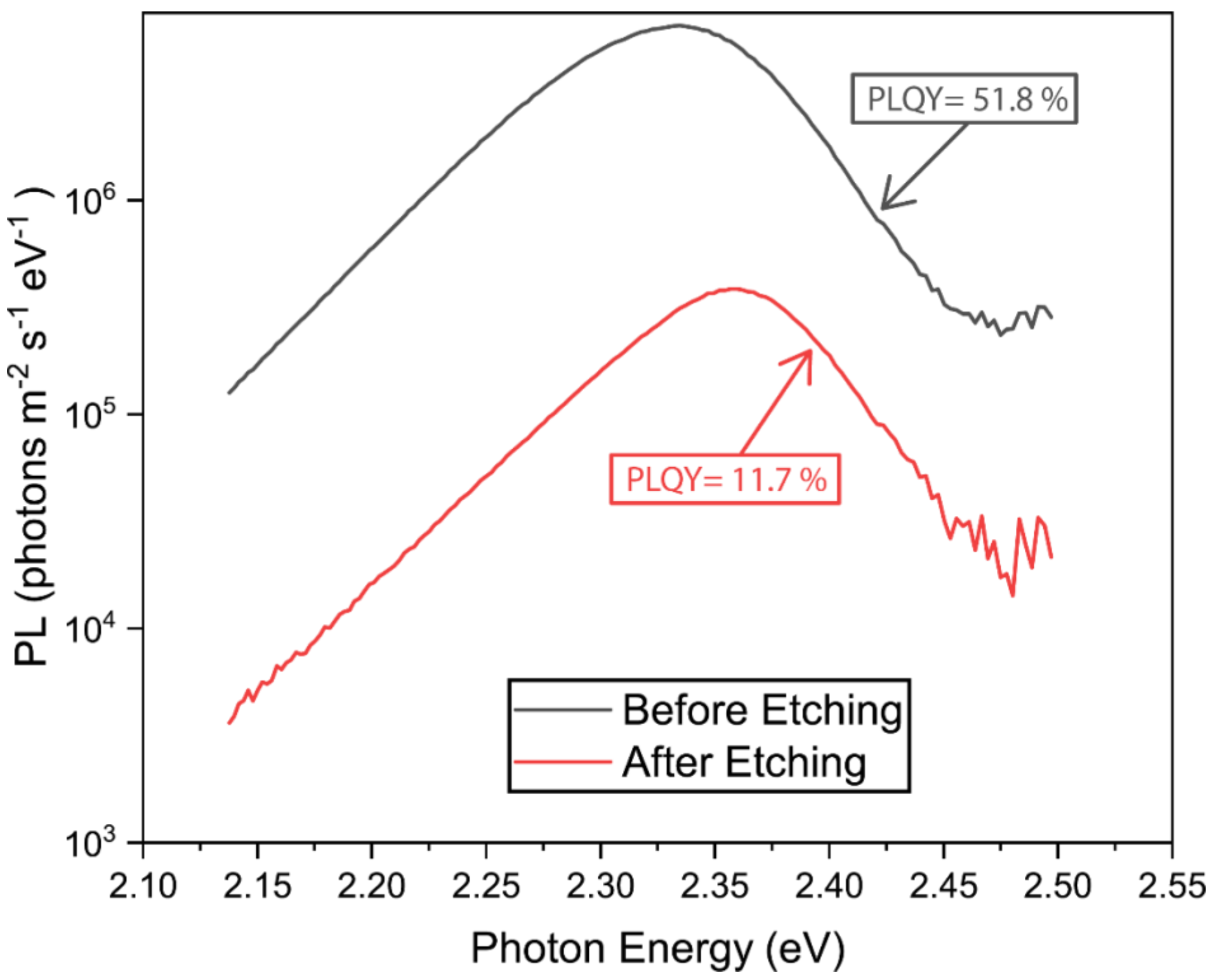

**Figure S15**. $FAPbBr_3$ NCs PLQY measurements before and after the etching process.